\documentclass[a4paper, prd, twocolumn, 
nofootinbib,superscriptaddress]{revtex4-2}

\usepackage{color}
\usepackage[usenames,dvipsnames]{xcolor}
\usepackage{multirow}
\usepackage{amstext}
\usepackage{amssymb}
\usepackage{amsmath}
\usepackage{mathrsfs}  
\usepackage[hidelinks]{hyperref}
\usepackage{paralist}
\usepackage{graphicx}
\usepackage{url}
\usepackage[normalem]{ulem} %
\usepackage{makecell} %

\graphicspath{{figs/}}

\def\d{{\mathrm{d}}}

\begin{document}

\title[]{Testing gravitational waveform models using angular momentum}

\author{Neev Khera
   }
\email{neevkhera@psu.edu}
\affiliation{Institute for Gravitation and the Cosmos, Pennsylvania State 
University, University Park, PA 16802, USA}

\author{Abhay Ashtekar
   }
\affiliation{Institute for Gravitation and the Cosmos, Pennsylvania State University, University Park, PA 16802, USA}

\author{Badri Krishnan
    }
\affiliation{Max Planck Institute for Gravitational Physics (Albert Einstein Institute), Callinstrasse 38, D-30167 Hannover, Germany}
\affiliation{Leibniz Universit\"at Hannover, 30167 Hannover, Germany}
\affiliation{Institute for Mathematics, Astrophysics and Particle Physics, Radboud University, Heyendaalseweg 135, 6525 AJ Nijmegen, The Netherlands}

\begin{abstract}

The anticipated enhancements in detector sensitivity and the corresponding increase in the number of gravitational wave detections will make it possible to estimate parameters of compact binaries with greater accuracy assuming general relativity(GR), and also to carry out sharper tests of GR itself. Crucial to these procedures are accurate gravitational waveform models. The systematic errors of the models must stay below statistical errors to prevent biases in parameter estimation and to carry out meaningful tests of GR. Comparisons of the models against numerical relativity waveforms provide an excellent measure of systematic errors. A complementary approach is to use  balance laws provided by Einstein's equations to measure faithfulness of a candidate waveform against exact GR. Each balance law focuses on a physical observable and measures the accuracy of the candidate waveform vis-\`a-vis that observable. Therefore, this analysis can provide new physical insights into sources of errors. In this paper we focus on the angular momentum balance law, using post-Newtonian theory to calculate the initial angular momentum, surrogate fits to obtain the remnant spin, and waveforms from models to calculate the flux. For brevity of presentation we restrict ourselves to the waveform models \texttt{IMRPhenomXPHM}, \texttt{NRSur7dq4},\texttt{SEOBNRv4PHM}, \texttt{IMRPhenomPv2}, and \texttt{SEOBNRv3}. The consistency check provided by the angular momentum balance law brings out the marked improvement in the passage from \texttt{IMRPhenomPv2} to \texttt{IMRPhenomXPHM} and from \texttt{SEOBNRv3} to \texttt{SEOBNRv4PHM} and shows that the most recent versions agree quite well with exact GR. For precessing systems, on the other hand, we find that there is room for further improvement, especially for the Phenom models. 
  
\end{abstract}

\maketitle

\section{Introduction}

\label{sec:intro}
The next generation of gravitational wave detectors with much higher 
sensitivity are on the horizon \cite{Punturo:2010zz, Reitze:2019iox, Audley:2017drz, Kawamura:2006up, Luo:2015ght}. We can expect detection of compact binaries with orders of magnitude higher signal-to-noise ratio than current measurements. Consequently it will allow unprecedented precision in the tests of general relativity in the highly-nonlinear regime. Moreover it will allow high-precision parameter estimation of the compact binary. However to carry out 
these procedures, it is essential to have accurate waveform models 
whose systematic errors are smaller than the measurement errors. 

Gravitational wave observations allow several families of tests of general relativity(GR) \cite{TheLIGOScientific:2016src, LIGOScientific:2019fpa, LIGOScientific:2020tif}. Many such tests can be done without waveform models, such  as parametrized tests of 
post-Newtonian (PN) theory \cite{Arun:2004hn, Arun:2006hn, Mishra:2010tp, Yunes:2009ke, Li:2011vx} ,tests with the quasinormal ringdown frequencies \cite{Isi:2019aib, Carullo:2019flw, Capano:2021etf}, or `no-hair' tests of the inspiral phase of binaries \cite{Dhanpal:2018ufk, Islam:2019dmk, Capano:2020dix}. However these tests rely on the analytic solutions from the perturbative regimes. For testing the highly nonlinear merger regime, waveform models are indispensable. For example one can perform the residual test, where the difference between the data and the best-fit waveform obtained from a model is tested for consistency with being purely noise\cite{LIGOScientific:2019fpa, LIGOScientific:2020tif}. Some tests can combine many events to have increasing stringency. However it has been shown that accuracy requirements of models also increase for such tests, and that current models may not be sufficiently accurate to perform such tests using detections made so far \cite{Moore:2021eok}.

Waveform models are created using a diverse range of innovative ideas. However to obtain any model it is necessary to make approximations, and the ensuing systematic errors are unavoidable. A useful way to measure the error is by computing the mismatch of the models against numerical relativity (NR) waveforms using a detectors noise spectrum. If the mismatch $\mathcal{M}$ between NR and the model satisfies $\mathcal{M}\leq 1/\rho^2$, where $\rho$ is the detector signal-to-noise ratio of an event, then the model will not have significant biases in parameter estimation \cite{Flanagan:1997kp, Lindblom:2008cm}. It has been argued that this sufficient condition can be relaxed in practice \cite{Chatziioannou:2017tdw}; nevertheless the mismatch must still scale as $1/\rho^2$. In these analyses one takes NR to be a proxy for the exact GR waveform. Therefore, the accuracy for NR must increase for future detectors as well \cite{Purrer:2019jcp}.

On the other hand there are additional tools to measure errors of waveform models from GR: Balance laws. The balance laws do not depend on NR and can thus be used at any point in parameter space, especially where NR simulations are sparse. Moreover, the balance laws may provide new insights into sources of errors. Exact GR in asymptotically flat spacetime has a large asymptotic symmetry group: the Bondi-Metzner-Sachs (BMS) group \cite{Bondi:1962px, Sachs:1962wk}. This group gives rise to infinitely many balance laws \cite{Ashtekar:1981bq, Dray:1984gz}. In addition to the more familiar energy, momentum, and the Poincar\'e angular momentum 
balance laws, there is an infinite family of supermomentum balance laws. Application of the supermomentum balance law to test waveform systematics was discussed in \cite{Ashtekar:2019viz,Khera:2020mcz}.  The application of the three-momentum balance laws has been discussed further more recently in \cite{Borchers:2021vyw}.

In this paper we will focus on using the angular momentum balance law. There is an important subtlety with angular momentum in asymptotically flat GR: The angular momentum suffers from an ambiguity, that arises from  supermomentum. However, a detailed analysis \cite{Ashtekar:2019rpv} has shown that this contribution leads to a correction term that is at most $\mathcal{O}(v^2)$ in compact binary coalescences, where $v$ is the kick velocity. Since this effect is too small for the level of accuracy of interest to this paper we will neglect it. 
Therefore, for our purposes, the angular momentum balance law can be stated simply as 
\begin{equation}
    \label{eq:balance_law}
    J^k(t_f) = J^k(t_i) + \mathcal{F}^k \, ,
\end{equation}
where $J^k$ denotes angular momentum vectors, $i=1,2,3$, and $t_i$ and $t_f$ are the initial and final times. Here $\mathcal{F}^k$ is the flux between $t_i$ and $t_f$, and can be expressed in terms of the gravitational strain $h^\circ = rh = r(h_+ - ih_\times)$, where $r$ is the luminosity distance to the source. We have,
\cite{Ashtekar:1981bq, Dray:1984gz}
 
\begin{align}
  \label{eq:flux}
  \mathcal{F}^k = \frac{i}{32\pi}\int_{t_i}^{t_f}\!\! \d t \, \d\Omega\, 
  \bar{\eth}\hat{r}^k\Big( \dot{h}^\circ \eth \bar{h}^\circ + 
  \dot{\bar{h}}^\circ \eth h^\circ &-  2\eth(\dot{h}^\circ\bar{h}^\circ) \Big)  \nonumber\\
  &+ \,\mathrm{c.c.}
\end{align}
with $\hat{r}^k = (\sin\theta \cos\phi, \sin\theta\sin\phi, \cos\theta)$. Note that since the flux depends only on $h^\circ$---rather than $h$---dependence of the balance law on the luminosity distance $r$ is factored out. For the definition of $\eth$  and all other conventions we follow the Moreschi-Boyle conventions (see Appendix B of \cite{Iozzo:2020jcu}). 

The idea is to test the accuracy of a candidate waveform vis-\`a-vis exact general relativity by checking how well it satisfies the balance law (\ref{eq:balance_law}). This requires us to evaluate each term in Eq.~\eqref{eq:balance_law}. The flux in Eq.~\eqref{eq:flux} can be evaluated directly from the strain given by the candidate waveform model. But to evaluate the initial and final angular momentum, one needs additional inputs. For the initial angular momentum we can resort to post-Newtonian theory, provided the initial time is chosen to be early enough that 
the PN expressions are sufficiently accurate. For the final spin, we will use fits to the dimensionless spin $\vec{\chi}_f$ and mass $M_f$ of the remnant black hole, provided by the surrogate fit \texttt{NRSur7dq4Remnant}  \cite{Varma:2019csw}.
However $\vec{\chi}_f$ gives just the intrinsic angular momentum of the remnant. Since we work in the rest frame of initial binary, generically the remnant is not at rest, whence the total final angular momentum has to be obtained by applying a boost to the intrinsic angular momentum. 
Nonetheless, the discrepancy between the intrinsic and total angular momentum is typically of order $10^{-5}$ \cite{Iozzo:2021vnq, Ashtekar:2019rpv}. This is smaller than the accuracy levels considered in this paper and thus we assume $J^k(t_f) = M_f^2 \chi_f^k$. Therefore we can calculate all the ingredients of Eq.~\eqref{eq:balance_law} and can evaluate the violation of the equality not only for different waveform models, but also for numerical simulations.

The plan for the rest of the paper is the following. In Sec.~\ref{sec:methodology} we shall expand on the various ingredients that are needed to test the angular momentum balance. Using these ingredients,  Sec.~\ref{sec:Results} tests the waveform models listed in Table~\ref{tab:models} as well as NR simulations. Finally Sec.~\ref{sec:discussion} concludes with a discussion of the results and possible future applications.  In Appendix~\ref{app:PN} we summarize the procedure used to calculate the initial angular momentum of the system using the 3.5 PN expansion of~\cite{Bohe:2012mr}. Appendix~\ref{app:Heat_map} provides plots and heat maps of the angular balance law violations by various models as functions of the binary parameters. These offer insights on the extent to which improvements in the effective one body (EOB) and Phenom models have decreased errors, and also provide guidance on regions of the parameter space where further improvements can be made. We use units with $G=c=1$.

\section{Methodology}
\label{sec:methodology}
To measure the violation of Eq.~\eqref{eq:balance_law} in a candidate waveform for binary black holes, we calculate the remnant dimensionless spin of the black hole using two techniques and compare them. As mentioned in the Introduction we work under the approximation
\cite{Iozzo:2021vnq} 
\begin{equation}
\label{eq:final_ang_mom}
    J^k(t_f) \approx M_f^2 \chi_f^k\,,
\end{equation}
where we have ignored terms $\sim 10^{-5}$ linear in kick velocity. Using Eq.~\eqref{eq:final_ang_mom},  the initial spin $J^k(t_i)$ provided by the PN expression, and the flux $\mathcal{F}$  calculated from Eq.~\eqref{eq:flux} using the candidate waveform, the balance law provides the final dimensionless spin $\vec{\chi}_{\mathrm{bal}}$:
\begin{equation}
    \label{eq:chi_bal}
    {\chi}^k_\mathrm{bal} = \frac{1}{M_f^2}\left(J^k(t_i) + \mathcal{F}^k\right)\,.
\end{equation}
On the other hand, we can also get the remnant dimensionless spin  $\vec{\chi}_\mathrm{fit}$ from the \texttt{NRSur7dq4Remnant} fit. Therefore by comparing $\vec{\chi}_\mathrm{fit}$ to $\vec{\chi}_\mathrm{bal}$ we can measure the deviation from the balance law in Eq.~\eqref{eq:balance_law}.

\subsection{Flux}
To calculate the flux of angular momentum in Eq.~\eqref{eq:flux}, we need the strain between an early time $t_i$ and a late time $t_f$. We can either use waveforms from NR, or from models. There is a wide variety of models, and we will use some state of the art models, as well as some older models for comparison. While several models have been left out for brevity of presentation, they can be analyzed using similar techniques. Ideas behind these models and details of how they are implemented can be found in the references; a discussion of this diverse material is beyond the scope of this paper. 

Three families of models have been extensively discussed in the literature. First are the EOB models \cite{Buonanno:1998gg,Bohe:2016gbl,Taracchini:2013rva,Pan:2013rra,Pan:2011gk,Nagar:2018zoe,Damour:2014sva, Pan:2013rra, Taracchini:2013rva, Babak:2016tgq, Ossokine:2020kjp}. See \cite{Damour:2016bks} for a review. The specific EOB models used in this paper are \texttt{SEOBNRv3} \cite{Pan:2013rra, Taracchini:2013rva, Babak:2016tgq} and \texttt{SEOBNRv4PHM} \cite{Ossokine:2020kjp}. The second family is the Phenom models
\cite{Ajith:2007kx, Khan:2019kot, Khan:2018fmp, London:2017bcn, Santamaria:2010yb, Husa:2015iqa, Khan:2015jqa, Hannam:2013oca} and  its cousin, the family of PhenomX models \cite{Pratten:2020ceb, Garcia-Quiros:2020qpx, Pratten:2020fqn}.
Specifically we use \texttt{IMRPhenomPv2}\cite{Husa:2015iqa, Khan:2015jqa, Hannam:2013oca} and \texttt{IMRPhenomXPHM} \cite{Pratten:2020ceb} models. Finally there is the family of surrogate waveform models \cite{Field:2013cfa, Blackman:2015pia, Blackman:2017dfb, Blackman:2017pcm, Varma:2019csw, Islam:2021mha}, 
from which we use the \texttt{NRSur7dq4}\cite{Varma:2019csw} model.
See \cite{Tiglio:2021ysj} for a review. 

All these waveform models first produce the strain in the coprecessing frame \cite{Boyle:2011gg}, and then apply a `twisting up' procedure to return the strain in the inertial frame. In the coprecessing frame different waveform models include different modes in their modeling. Omission of modes can introduce significant modeling errors to the flux. Table~\ref{tab:models} shows the lists of modes included in the waveform models considered in this paper.
\begin{table}
    \centering
    \begin{tabular}{l@{\hspace*{4em}}c}
        \Xhline{3\arrayrulewidth}
        Waveform model &  Coprecessing modes included \\ \hline
         \texttt{SEOBNRv3} & $(2,\pm2)$, $(2,\pm1)$ \vspace*{0.25em}\\
         \texttt{SEOBNRv4PHM}& $(2,\pm2)$, $(2,\pm1)$, \\ 
                            & $(3,\pm3)$, $(4,\pm4)$, $(5,\pm5)$  \vspace*{0.25em}\\
         \texttt{IMRPhenomPv2}& $(2,\pm2)$  \vspace*{0.3em} \\
         \texttt{IMRPhenomXPHM}& $(2,\pm2)$, $(2,\pm1)$, \\
                                &$(3,\pm3)$,  $(3,\pm2)$,  $(4,\pm4)$ \vspace*{0.25em}\\
         \texttt{NRSur7dq4}& All modes with $\ell\leq4$ \\
         \Xhline{3\arrayrulewidth}
    \end{tabular}
    \caption{The waveform models used in the paper and the modes of the waveform they include in the coprecessing frame of the binary black hole. For precessing systems the coprecessing frame is `twisted up' into the inertial frame to obtain the final waveform. }
    \label{tab:models}
\end{table}

Once we obtain the strain, to evaluate Eq.~\eqref{eq:flux} we find it useful to expand the strain in terms of the spin-weighted spherical harmonics. For models that do not provide the mode decomposition by default, specifically the Phenom models, we evaluate the waveform on a grid across the sky and transform into the spin-weighted spherical harmonic basis using the code \texttt{spinsfast}\cite{Huffenberger:2010hh}. The spin-weighted spherical harmonics are eigenvectors of the angular derivative $\eth$, simplifying the calculations. We also express $\hat{r}^k$ in terms of spherical harmonics. Then the integrand of Eq.~\eqref{eq:flux} turns into a product of spin weighted spherical harmonics, and the angular integration can be evaluated using the formula for the integral of their triple product.  Finally we perform the time integral numerically to obtain the flux.

\subsection{Post-Newtonian angular momentum}
\label{subsec:PN}
Critical to our analysis is an expression of the initial angular momentum of the system. To obtain it, we resort to post-Newtonian theory, see \cite{Blanchet:2013haa, Schafer:2018kuf, Porto:2016pyg} for reviews. The total angular momentum $J^k$ is traditionally split into the orbital and spin angular momentum, $L^k$ and $S^k$ respectively, 
\begin{align}
    J^k = L^k + S^k,
\end{align}
with $S^k = m_1^2 \chi_1^k + m_2^2 \chi_2^k$. Here we are interested in the center of mass frame description of binary black holes that are in quasicircular orbits. Let the orbital frequency of the binary at the initial time be $\Omega_{\mathrm{orb}}$. For quasicircular orbits we can expand PN expressions in the gauge invariant dimensionless  parameter $x=(G M \Omega_{\mathrm{orb}})^{2/3}$. This parameter allows us to connect with the waveform models, where their start times are specified in terms of $\Omega_\mathrm{orb}$ or the frequency of the coprecessing $(2,2)$ mode, $f_{22} \approx \Omega_\mathrm{orb}/\pi$ \cite{Boyle:2013nka}. Thus we would like a PN expression of the orbital angular momentum $L^k(m_1,m_2, \vec{\chi}_1, \vec{\chi}_2, x)$ for a quasicircular binary in the center of mass frame. However the spin-spin interaction terms starting at 2 PN order cause quasicircular orbits to radially oscillate at the orbital timescale, complicating calculations \cite{Bohe:2015ana}.  (For nonprecessing systems these issues do not arise and angular momentum can be calculated with spin-spin terms as well \cite{Bohe:2015ana, Porto:2010zg, Cho:2021mqw}.) If only linear in spin terms are kept, an expression for the angular momentum in the desired form is given  in Eq.~(4.7) of \cite{Bohe:2012mr} up to 3.5PN. We use this expression for the angular momentum in our work. (See the Appendix A for details on how we apply the PN formula.) Thus, we make two approximations in the calculation of $J^k(t_i)$: truncating of the PN expansion at 3.5 order and ignoring all the nonlinear spin contributions.
While we do not estimate systematic errors due to nonlinear spin interactions, as is common in the literature, we estimate the truncation errors by comparing the 3.5 PN and 3 PN results. 
\subsection{Remnant angular momentum}
\label{subsec:remnant}
To obtain the final angular momentum we use fits to NR values of the remnant final mass and dimensionless spin. In NR the values are typically calculated using quasilocal measures \cite{Ashtekar:2003hk} on the horizon. But it has also been calculated asymptotically 
\cite{Iozzo:2021vnq}  for 13 simulations and agreement with the horizon values is excellent. In our analysis we use the horizon values since they are reported by all simulations. These values can then be interpolated across parameter space using a catalog of numerical simulations. While there are several such fits in the literature, we use the \texttt{NRSur7dq4Remnant} \cite{Varma:2018aht, Varma:2019csw}  fit. This fit provides us with the remnant mass and the dimensionless spin vector (as opposed to just the magnitude), as well as an estimate of their respective errors.%
\footnote{Note that the calculation of  $\vec{\chi}_{\rm fit}$ in \texttt{NRSur7dq4Remnant} is \emph{independent} of
 the waveform model \texttt{NRSur7dq4}, and thus the satisfaction of the balance law for the surrogate wave form is not tautological; it is a nontrivial consistency check.} 
Thus we are able to obtain the remnant angular momentum using Eq.~\eqref{eq:final_ang_mom}.

There is a subtlety with using the fit for precessing systems: The spins precesses over time, thus the same system can be labeled by different spins at different reference times. For various applications it is useful to have the ability to use the fit with the spins specified at arbitrary times. This is done in \texttt{NRSur7dq4Remnant} by implementing a model for the spin evolution to evolve the spins to a standard time of $100M$ before the peak of the strain, where $M$ in the total mass of the system. The spins at $100M$ are now used for the interpolation of the remnant quantities.  However the spin evolution introduces new errors to the fit. These errors are harder to estimate and are not accurately provided by the model, as already noted when the fit was introduced in \cite{Varma:2019csw}. Thus the error estimates returned by \texttt{NRSur7dq4Remnant} must be taken with a grain of salt, as it only represents the errors from the Gaussian process regression procedure used by the model.

In Sec.~\ref{sec:Results}  we will find that these estimates are too small compared to
the actual errors for precessing systems, as calculated from comparisons to NR. So instead of using the error estimates from the fit, we use the comparison with NR to provide us with an estimate of the errors involved.

\section{Results}
\label{sec:Results}

We now apply the methods discussed to waveform models
as well as to NR simulations. To test the waveform models across parameter space we select random points in parameter space and check violations of the balance law. We divide our study of the models in two parts: precessing and nonprecessing systems. For both these families we restrict the parameter space to a finite compact region. Since we are dealing with binary black holes that are initially in  quasicircular orbits, the parameter space is described by the mass ratio $q$ and the dimensionless spins $\vec{\chi}_1, \vec{\chi}_2$. We restrict these parameters to be within range of applicability of \texttt{NRSur7dq4}. Additionally, since \texttt{NRSur7dq4} only models waveforms for finite time,  we would like the \texttt{NRSur7dq4} waveforms to be long enough so that we can use PN methods at its start. While \texttt{NRSur7dq4} goes up to mass ratio $4$, the waveforms start at higher frequencies with increasing mass ratio. Therefore to be able to safely use PN expressions, initially we restrict the mass ratio to $1\leq q\leq 2$. This allows us to safely use waveforms starting at $5.8\times10^{-3}$ in dimensionless units. Additionally we also restrict spin magnitudes to be less than $0.8$ to be within the training data range of \texttt{NRSur7dq4}, as well as the remnant data fit \texttt{NRSur7dq4Remnant} that we use. 

For the NR simulations we use the publicly available SXS catalog \cite{Boyle:2019kee} of NR simulations. But we restrict consideration to numerical simulations that lie in the parameter range considered above.

\subsection{Nonprecessing systems}
\label{subsec:aligned}
In this section we test satisfaction of the balance law for randomly selected 20,000 non-precessing points in the parameter space. The spins are in the $z$-direction with $\chi_1^z$ and $\chi_2^z$ uniformly and independently distributed in the interval $[-0.8,0.8]$. 
We obtain the distribution of mass ratio $q$ indirectly from the distribution of masses $m_1$ and $m_2$ to replicate commonly chosen priors. We take masses $m_1$ and $m_2$ to be independent and uniform, subject to constraints $q<2$ and $20<m_1+m_2<160$. For later convenience we would like $m_1\geq m_2$, therefore if $m_1<m_2$we exchange the labels. Then for each of these points, we will test how well the balance law is satisfied.

\begin{figure}[tb]
    \centering
    \includegraphics{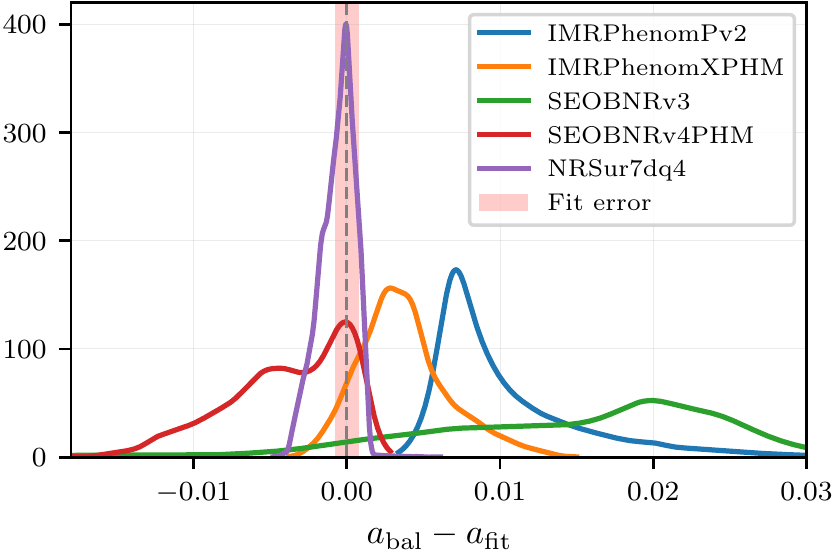}
    \caption{Non-precessing systems: The distribution of the difference $(a_{\mathrm{bal}} - a_{\mathrm{fit}})$ between the magnitudes of the remnant spin calculated by using the angular momentum balance law and using the fit \texttt{NRSur7dq4Remnant}.
    The distribution is calculated for different waveform models using the same sample points. The shaded region shows the error estimate of the fit. 
    }
    \label{fig:NP_spin}
\end{figure}

We first calculate the spin of the remnant black hole $\vec{\chi}_{\text{bal}}$ using the balance law,  from Eq.~\eqref{eq:chi_bal}. For non-precessing systems, by symmetry we have that $\vec{\chi}_{\text{bal}} = a_{\text{bal}} \hat{z}$. We can compare this to the remnant spin $\vec{\chi}_{\text{fit}}=a_{\text{fit}}\hat{z}$ obtained from the fit \texttt{NRSur7dq4Remnant}. Mismatch between $\chi_{\rm bal}$ and $\chi_{\rm fit}$ provides us the desired measure of accuracy of the waveform model under consideration.  In Fig.~\ref{fig:NP_spin} we plot the distribution of $a_{\text{bal}} - a_{\text{fit}}$ across the random points in parameter space. To help identify the errors coming from waveform modeling, we also show an estimate of the errors from the fit, obtained from \texttt{NRSur7dq4Remnant}. For the region of the parameter space considered, the error estimates are almost constant (the standard deviation of the distribution of errors being  $~1\%$ of the mean). We thus use the mean of this distribution as the fit error for our samples.  Although the PN truncation error is not shown in the plot, it is $65\%$ of the fit error, but it does not include the errors from ignoring spin-spin interaction terms.

Fig.~\ref{fig:NP_spin} shows that, overall, the agreement between $a_{\rm bal}$ and $a_{\rm fit}$ is of order $10^{-2}$. Moreover we see clear evidence for the improvement of \texttt{SEOBNRv4PHM} over \texttt{SEOBNRv3} and of \texttt{IMRPhenomXPHM} over \texttt{IMRPhenomPv2}.  The surrogate model has the best performance, with all the balance law violation consistent with solely coming from the fit and PN truncation errors. By comparison, although the mismatch is only at a $10^{-2}$ level for
EOB and Phenom, the modeling errors are significantly larger than those coming from the fit and PN truncation errors; thus there is room for further improvement.

\begin{figure}[tb]
    \centering
    \includegraphics{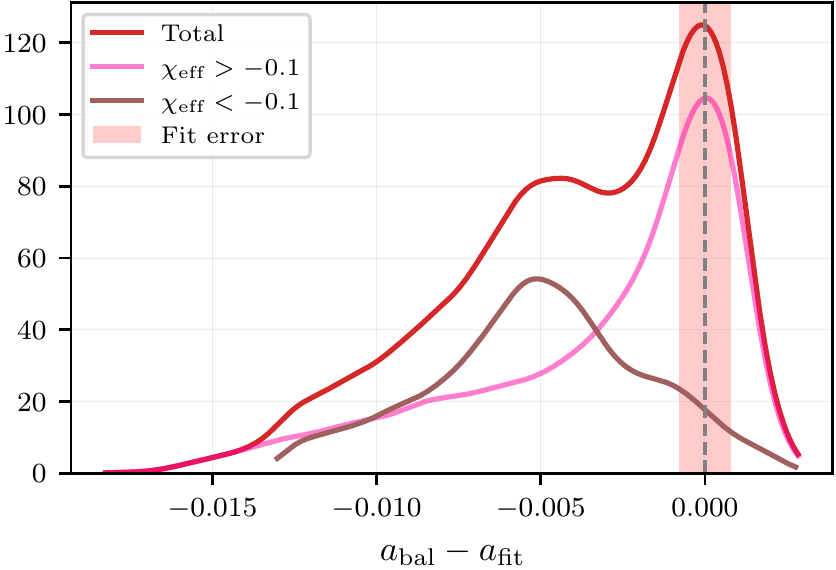}
    \caption{The distribution of balance law violation for \texttt{SEOBNRv4PHM} from Fig.~\ref{fig:NP_spin}. Here we have split the points in parameter space in two, with $\chi_{\mathrm{eff}}<-0.1$ and $\chi_\mathrm{eff}>-0.1$. This split separates the double hump in \texttt{SEOBNRv4PHM}, and shows us that the balance law violation is larger for negative $\chi_\mathrm{eff}$.}
    \label{fig:EOB_chi_eff}
\end{figure}

Note also that for \texttt{SEOBNRv4PHM} the plot has an interesting double hump. We find that these humps are correlated with the effective spin parameter $\chi_\mathrm{eff}$ defined as 
\begin{align}
    \label{eq:chi_eff}
    \chi_\mathrm{eff} = \frac{m_1\chi_1^z + m_2\chi_2^z}{m_1+m_2} \, .
\end{align}
The correlation---shown in Fig.~\ref{fig:EOB_chi_eff}---brings out the sharp difference between distributions for $\chi_\mathrm{eff}<-0.1$ and $\chi_\mathrm{eff}>-0.1$, as can be seen in more detail from the dependence of the balance-law violation on $\chi_\mathrm{eff}$ (See Appendix~\ref{app:Heat_map}). This illustrates the power of the balance law to identify regions of parameter space where errors are higher, thereby providing guidance for further improvements of the waveform model. 

\subsection{Precessing systems}
\label{subsec:precessing}
As in Sec.~\ref{subsec:aligned}, we randomly select 20,000 points in parameter space, but now using precessing systems, and evaluate the violation of the angular momentum balance law for them. The spins are sampled independently with an isotropic distribution. The spin magnitude is taken to be uniformly distributed in $[0,0.8]$. The mass ratio is sampled from the same distribution as in Sec.~\ref{subsec:aligned}.

\begin{figure}[tb]
    \centering
    \includegraphics{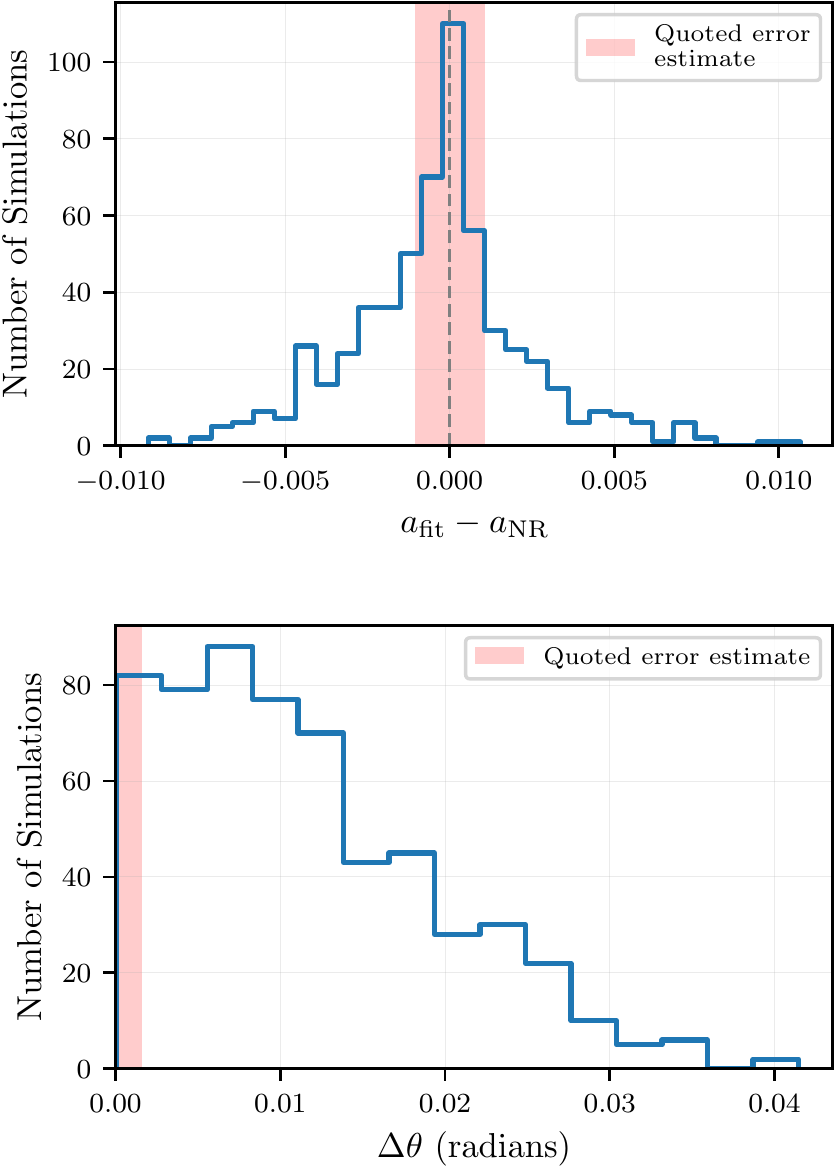}
    \caption{Comparison of the remnant spin from 672 precessing NR simulations that lie in the parameter range and starting frequency considered in the paper, to the fit \texttt{NRSur7dq4Remnant}. The shaded region shows the error estimate provided by the fit model. However as noted in \cite{Varma:2019csw}, this estimate does not include errors from the spin evolution; it only includes errors from the Gaussian process regression procedure that is used. The upper plot shows the difference in the magnitude of spins, and the lower plot shows the angle between them. We see that for the parameters we consider and for the starting frequency we use, the real errors are much larger than the estimates. We use error estimates obtained from these 672 NR simulations for the rest of the paper.}
    \label{fig:fit_vs_NR}
\end{figure}

The remnant spin is now arbitrarily oriented. Therefore, to compare $\vec{\chi}_{\text{bal}}$ with $\vec{\chi}_{\text{fit}}$ we are led to compare their magnitudes $a_{\text{bal}}$ and $a_{\rm fit}$, and also to calculate the angle $\Delta\theta$ between them. However
there is a difference in the calculation of error estimates because,
as discussed in Sec.~\ref{subsec:remnant}, for precessing systems the fitting procedure complicated by evolution of spin with time. This is accounted for by using a spin evolution model, which introduces further errors in $a_{\rm fit}$ and $\Delta\theta$. The reported error estimates from the fit \texttt{NRSur7dq4Remnant} do not include these errors. Therefore we will estimate these errors by a direct comparison with NR simulations. 
The NR simulations are taken from the SXS public catalog \cite{Boyle:2019kee} of NR simulations. We choose quasicircular binary black hole simulations that are long enough to include our choice of starting frequency and have parameters that lie within the range under consideration in this paper. We also drop the first 337 older simulations, and we are then left with 672 precessing NR simulations. For these simulations we compute the remnant spin using the fit and compare to the actual NR value. The result is shown in Fig.~\ref{fig:fit_vs_NR}, where we see that the error quoted in \texttt{NRSur7dq4Remnant} is much smaller than the actual error. We thus compare the balance law violations with the errors from these simulations instead of using the error from the fit. However because the fit is trained against these simulation, the errors might in fact be larger for regions of parameter space with a scarcity of simulations and moreover these 672 simulations do not represent an unbiased sampling of the parameter space considered here. Nonetheless for the rest of this paper we use this error distribution, keeping in mind that they are not meant to be sharp.  

\begin{figure}[tb]
    \centering
    \includegraphics{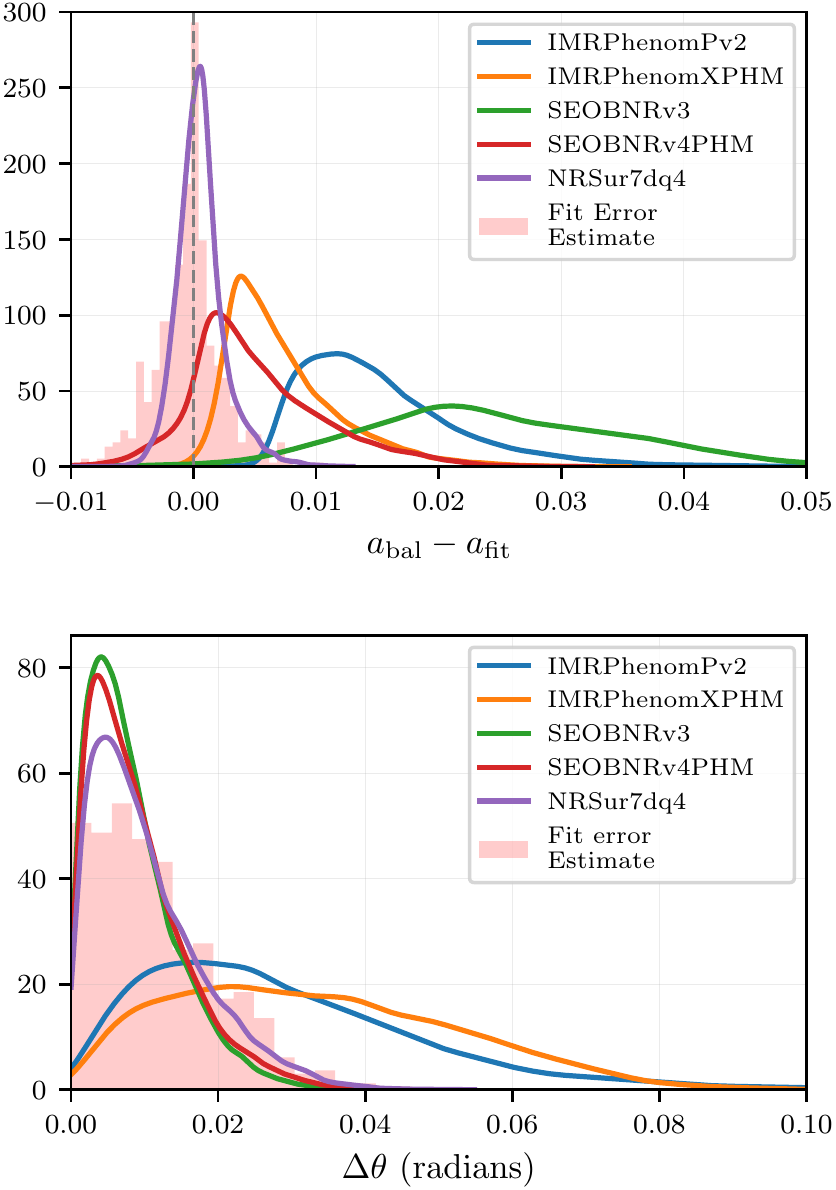}
    \caption{ Precessing systems: The distribution of angular momentum balance-law violation across the parameter range considered in the paper, using various waveform models. The upper plot shows the difference between the magnitudes of the remnant spin $a_{\mathrm{bal}}$, computed from the balance-law, and $a_{\mathrm{fit}}$, computed using the fit \texttt{NRSur7dq4Remnant}. The lower plot shows the angle $\Delta\theta$ between the remnant spin computed using the two different methods. We also show in the shaded region the distribution of fit errors estimated from direct comparison with NR as in Fig.~\ref{fig:fit_vs_NR}}, as opposed to the quoted error estimate in the fit. 
    \label{fig:Prec}
\end{figure}

Using the error estimates discussed above, let us examine the violations of the angular momentum balance law. In Fig.~\ref{fig:Prec} we see the waveform models continue to have errors of order $10^{-2}$, albeit with larger errors than in the non-precessing case. For comparisons of the magnitude of the remnant spin,  \texttt{NRSur7dq4} again has the best performance, and its balance-law violations are consistent with the fit errors from NR.  The PN truncation error is only $9\%$ of the $90\%$ interval of the fit error here. The accuracy of the latest EOB and Phenom models, \texttt{SEOBNRv4PHM} and \texttt{IMRPhenomXPHM}, are very similar to each other. Furthermore,  we can clearly see the improvement of these EOB and Phenom models over their older versions. On the other hand, we see different results for the error in the angle in the lower plot. The lower plot of Fig.~\ref{fig:Prec} shows that the surrogate and EOB models have violations of the direction of spin that are consistent with the fit errors. The PN truncation error is negligible, only $0.7\%$ of the fit error. However the Phenom models show violations in the angle that are much larger than the errors. Thus, our analysis again provides pointers for further improvement. 

Finally, we also examine how the balance-law violation seen here varies with parameter space. This allows us to examine regions of parameter space where different models have deficiencies and provides guidance on improving this for future models. See Appendix~\ref{app:Heat_map} for the analysis.

\subsection{Lessons from and for NR}
\label{app:NR}

We now apply the angular momentum balance-law directly to NR simulations and discuss its implications. The procedure is almost identical to the one we used for waveform models, but uses the NR waveform instead of the model waveform.  More precisely, each NR simulation provides us with the waveform to calculate the flux $\vec{\mathcal{F}}$. The waveforms are known to improve significantly in accuracy by adding the memory effect \cite{Mitman:2020bjf, Talbot:2018sgr}. Therefore we add this effect to the waveforms to improve the accuracy of the calculated flux. The waveforms are labeled by the masses, spins, orbital frequency and separation of the two  progenitors at the starting time. Using these parameters and the 3.5 PN truncation discussed in Sec.~\ref{subsec:PN}, we calculate the initial angular momentum $\vec{J}(t_i)$ that is needed in Eq.~(\ref{eq:chi_bal}) of $\vec{\chi}_{\rm bal}$. For the remnant spin $\vec{\chi}_{{}_\mathrm{NR}}$, however, there is a key difference. We do not need the fit since we can directly use the remnant spin computed in the NR simulation at the horizon. The difference $\vec{\chi}_{\rm bal} - \vec{\chi}_{{}_\mathrm{NR}}$ measures the violation of the balance-law. There is, however, a subtlety: Since the binary system in NR may not be in the same reference frame in numerical simulations as in the frame we use for the  PN expression, we must perform a rotation to match the frames. For details see Appendix~\ref{app:PN}.  

\begin{figure}
    \centering
    \includegraphics{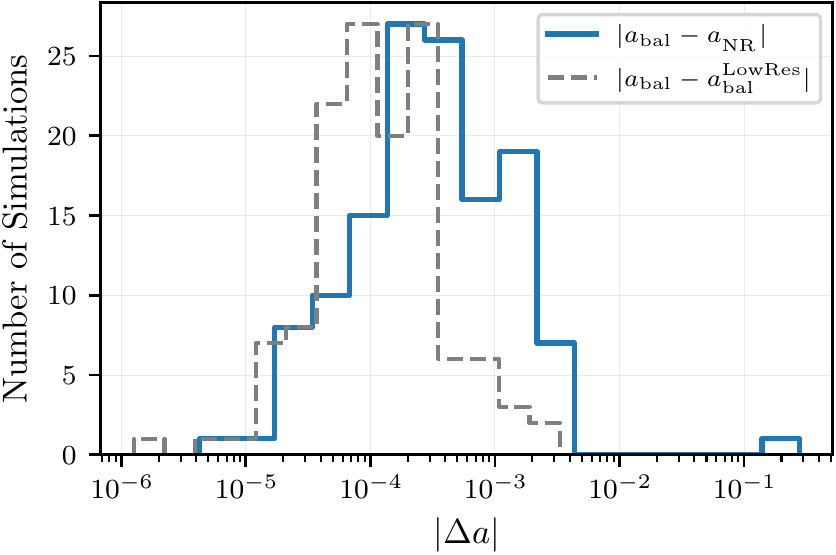}
    \caption{The violation of angular momentum balance-law for the 131 non-precessing numerical simulations described in the text. The solid blue curve shows the difference $a_\mathrm{bal} - a_{{}_\mathrm{NR}}$ between the magnitudes of the remnant spin  computed using the balance-law, and of the horizon spin.  The dashed gray line represents the numerical convergence error, i.e., the difference between the spin magnitudes, $a_\mathrm{bal}$ and $a_\mathrm{bal}^\mathrm{LowRes}$, computed using the highest and a lower resolution NR simulation.}  
    \label{fig:NR_mag_NP}
\end{figure}

We use the subset of simulations from the SXS public catalog \cite{Boyle:2019kee} described in Sec.~\ref{subsec:precessing}. However we further restrict ourselves to simulations where a lower resolution run is included, allowing us to analyze numerical errors. There are 131 such non-precessing NR simulations and 550 such precessing simulations. For all these simulations we calculate the remnant spin $\vec{\chi}_{\mathrm{bal}}$ from Eq.~\eqref{eq:chi_bal} with the highest resolution run available. Then we take the second highest resolution waveform to compute $\vec{\chi}_\mathrm{bal}^\mathrm{LowRes}$. Finally, by comparing $\vec{\chi}_\mathrm{bal}$ to  $\vec{\chi}_\mathrm{bal}^\mathrm{LowRes}$ we obtain an estimate of the numerical convergence errors, and by comparing $\vec{\chi}_\mathrm{bal}$ to the horizon spin $\vec{\chi}_{{}_{\mathrm{NR}}}$ we obtain a quantitative measure of the violation of the  balance-law.

In Fig.~\ref{fig:NR_mag_NP} the solid (blue) curve shows the violation of the angular momentum balance-law for the non-precessing simulations.   While the limited number of simulations makes a direct comparison with Fig.~\ref{fig:NP_spin} difficult, it is clear that overall the errors are manifestly smaller. However there is one outlier simulation \texttt{SXS:BBH:1134} with an error of order $10^{-1}$. On closer inspection we found that the orbital frequency is erroneous in the metadata file for that simulation, and computing the orbital frequency using the code \texttt{scri}\cite{scri, Boyle:2013nka, Boyle:2011gg, Boyle:2015nqa} from the waveform, $\lvert a_\mathrm{bal} - a_\mathrm{NR} \rvert$ is brought down to $\sim 1.5\times10^{-3}$ from $\sim 0.2$.  This is a concrete illustration of checks that balance-law considerations can provide on NR simulations themselves.

Aside from the outlier, we also see that the numerical errors are too small to account for the level of violation of the balance-law shown in Fig. \ref{fig:NR_mag_NP}. We also find that the PN truncation error obtained by comparing 3.5 PN to 3 PN is less than $3.5\times10^{-4}$ for all these simulations,  which is insufficient to account for the violation we found. What then is the main source of the violation? While in principle this discrepancy could be due to systematic errors in NR, it is much more likely that its origin lies  primarily in the assumption that nonlinear spin-spin interaction terms can be neglected in the PN calculation of $\vec{J} (t_{i})$.

\begin{figure}
    \centering
    \includegraphics{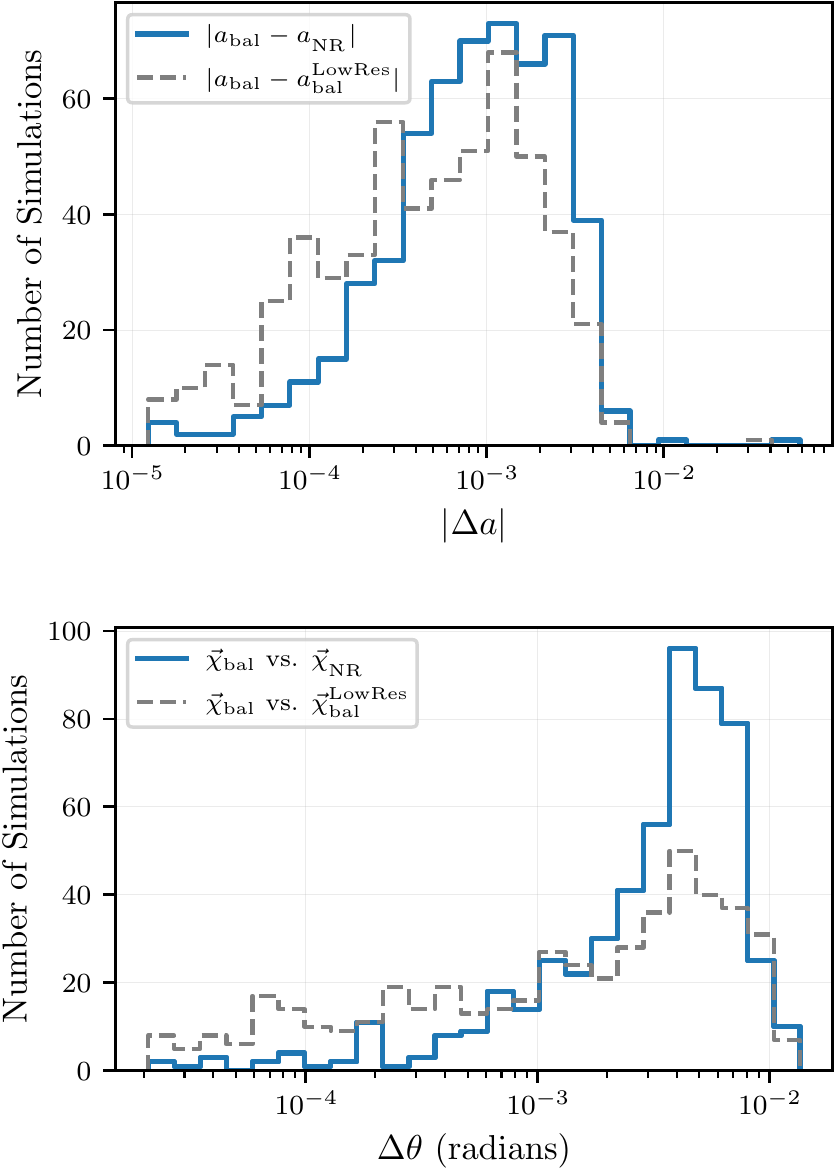}
    \caption{The angular momentum balance-law violation for the 550 precessing numerical simulations described in the text. The upper plot shows the violation in the magnitude of the spin, and is the same as Fig.~\ref{fig:NR_mag_NP} but for the precessing simulations. The lower plot in the solid blue line shows the angle $\Delta\theta$ between $\vec{\chi}_\mathrm{bal}$, the remnant dimensionless spin computed using the balance-law, and $\vec{\chi}_{{}_\mathrm{NR}}$ computed from the horizon. The dashed gray line represents the numerical convergence error, and is computed as the angle $\Delta\theta$ between  $\vec{\chi}_\mathrm{bal}$ and the same quantity computed using a lower resolution numerical simulation, $\vec{\chi}_\mathrm{bal}^\mathrm{LowRes}$.   }
    \label{fig:NR_mag_ang_P}
\end{figure}

Figure~\ref{fig:NR_mag_ang_P} shows the degree of violation of balance-law---as measured by the mismatch of $a_{\rm bal}$ and $a_{{}_{\rm NR}}$, and by the angle between $\vec{\chi}_{\rm bal}$ and $\vec{\chi}_{{}_{\rm NR}}$---as well as the convergence error for precessing systems.
Overall, the numerical convergence errors are larger than those in the nonprecessing case shown in Fig.~\ref{fig:NR_mag_NP} and match the scale of balance-law violations. However several individual simulations still have balance-law violations much higher than their respective numerical errors. The PN truncation error obtained by comparing 3.5PN to 3PN is less than $3.8\times 10^{-4}$ for the magnitude and $3.7\times 10^{-4}$ for the angle for all simulations. We also see an outlier simulation in the upper plot of Fig.~\ref{fig:NR_mag_ang_P}. This is the run \texttt{SXS:BBH:1131}. Unlike the previous outlier, we were not able to identify why the error is high nor were we able to ascertain anything special about the parameters. Therefore this simulation warrants attention of the NR community.

There are also lessons from NR simulations. That the violations of the balance-law in NR simulations are so small provides considerable confidence in the overall procedure. Furthermore, the remaining discrepancies provide a useful bound on the errors that come from the underlying assumptions and approximations. Notably we learn that the nonlinear spin-spin interaction terms that have been ignored in the PN angular momentum calculation can indeed be neglected at the current accuracy level of the waveform models. Secondly, the implicit assumption about the correspondence of PN and NR parameters is also tested here. The masses and spins, and especially the direction of the spin, are defined using distinct procedures in NR and PN. The direction of the spin is in fact not even a gauge invariant quantity in PN \cite{Bohe:2012mr} or NR\cite{Owen:2017yaj}. Therefore, a priori we do not have a reliable estimate on the discrepancies between the NR and PN assignments of these parameters. Again, the accuracy to which the balance-law is satisfied serves to provide assurance that the discrepancies are small for the level of accuracy of the current waveform models.

\section{Discussion}
\label{sec:discussion}

Different observables in full, nonlinear general relativity can be used to test different aspects of the accuracy of candidate waveforms. In this paper we focused on the angular momentum of black hole binaries. The angular momentum balance-law brings together diverse ideas:  post-Newtonian theory, numerical relativity, waveform modeling and calculations of the mass and spin of the remnant using surrogate fits. It is rather remarkable that all these ingredients come together in a consistent and precise manner. This overall coherence provides us some nontrivial checks. For example, we found that the spins measured from the horizon in NR matches very closely to the PN definition of spin, even though the direction of this three-vector is gauge dependent in the PN and NR analysis. We also found that the waveform models capture the physics of radiated angular momentum quite well as the system evolves from the inspiral, to the merger, and then ringdown, although some models capture it better than others.   

The balance-law provides us a new measure to test the accuracy of binary black hole waveform models, complementary to comparisons against NR. It allows us to not only compare the performance of different models, but identify regions of parameter space where errors are large \emph{without} directly using numerical simulations. We use PN methods to get the initial angular momentum, and the waveform to obtain the flux of angular momentum. Then the balance-law in Eq.~\ref{eq:chi_bal} gives us the remnant spin $\vec{\chi}_{\text{bal}}$. (Here we ignored the kick velocity and supertranslation corrections as they are much smaller than the level of accuracy of interest to this paper.) We then compared this $\vec{\chi}_{\text{bal}}$ to the spin $\vec{\chi}_{\mathrm{fit}}$ obtained by the remnant fit \texttt{NRSur7dq4Remnant} (which is conceptually independent from the waveform model \texttt{NRSur7dq4}). 

We first applied this procedure to waveform models with nonprecessing parameters and presented the results in Fig.~\ref{fig:NP_spin}. We found that the surrogate model \texttt{NRSur7dq4} performs exceptionally well, in that the modeling errors are at most the same order as errors from PN or from the fit to the remnant spin. The \texttt{SEOBNRv4PHM} and \texttt{IMRPhenomXPHM} models are close behind. The balance-law test also provided a sharp measure of the improvements over the older EOB and Phenom models, in part because, as Table \ref{tab:models} shows, they incorporate modes that their previous versions did not. Finally we also found that \texttt{SEOBNRv4PHM} has higher errors for parameters that correspond to negative effective spin $\chi_{\mathrm{eff}}$, as illustrated in Fig.~\ref{fig:EOB_chi_eff}. This difference illustrates the utility of using the balance-law to identify regions of parameter space with higher errors, on which efforts for future improvements could focus. A more fine-grained study could reveal more such regions.

For precessing systems, as discussed in Sec.~\ref{subsec:precessing}, the remnant fit \texttt{NRSur7dq4Remnant} has to model the spin evolution of the individual black holes. It was  noted in \cite{Varma:2019csw} that this evolution code introduces new errors. These errors are difficult to estimate accurately and were not included in \texttt{NRSur7dq4Remnant}, thus only the errors from the Gaussian process regression used in the procedure is reported. As Fig.~\ref{fig:fit_vs_NR} shows, for the typical spin evolution in our parameter space the Gaussian process regression errors are much smaller than the `real errors',  obtained by comparison with NR. We used the comparison to NR simulations to get a better error estimate for precessing systems. However, as we emphasized in Sec.~\ref{subsec:precessing}, this estimate is not as precise as it is for the nonprecessing systems. 

With this caveat in mind, we applied the balance-law to a distribution sample points in the parameter space describing precessing systems. For these systems the remnant spin need not be along the $z$-axis. Therefore,  we could measure the violations in the magnitude of angular momentum, as well as the direction. As seen in Fig.~\ref{fig:Prec}, for errors in magnitude we found that \texttt{NRSur7dq4} again has the best performance, with violations within the error scale. However, \texttt{SEOBNRv4PHM} and \texttt{IMRPhenomXPHM} are not far behind, and are very close to each other. They also showed clear improvements over their older versions. The situation turned out to be quite different for errors in angle. The fitting errors in the angle are large, but the surrogate and EOB models show violations only within the scale of this error and their predictions are almost identical to each other. So, for these models the fitting errors dominate and these models pass the balance-law test within the accuracy we can consistently demand. However the Phenom models perform poorly in comparison, and there has been no improvement over its older version. This suggests that there is room for improvement. Since Phenom performs well for nonprecessing systems, it seems likely that the likely culprit is the twisting up procedure used in this model.

In Appendix~\ref{app:Heat_map} we also analyze how the balance-law violations vary with parameter space for both precessing and nonprecessing systems considered above. This analysis provides insights into the regions of parameter space where errors are large and can help make improvements for future models.

We also applied the the balance-law to NR simulations. As one would expect, the simulations  perform better than the models. The high accuracy to which the balance-law is satisfied provides considerable confidence in the overall procedure, including the use of the approximation in which nonlinear spin-spin interaction terms are neglected. However, we also found that numerical convergence errors do not by themselves account for the violation of the balance-law. Thus, there is room to improve the accuracy of the additional ingredients that went into the procedure. 
Finally, the use of the balance-law enabled us to find two outliers in the NR simulations. We were able to identify the underlying problem in the first, \texttt{SXS:BBH:1134}, as having faulty metadata. However we do not know why \texttt{SXS:BBH:1131} has significantly larger errors; we hope it will receive further scrutiny from the NR community.

This work can be extended in several ways. In this paper we restricted the parameter space under consideration to include \texttt{NRSur7dq4} in the analysis.  While the model can go up to mass ratio $4$, the starting frequency becomes higher with increasing mass ratio, and the use of PN results becomes less reliable. However the remnant fit \texttt{NRSur7dq4Remnant} can be used with any starting frequency. Thus if we use only the EOB and Phenom models, the analysis can be extended to higher mass ratios. The starting frequency can also be lowered to reduce PN truncation effects. This enlarged parameter space has fewer NR simulations, and thus it would be interesting to identify regions where the models perform poorly using the balance-law. Furthermore although the remnant fits can be used to spin magnitudes above $0.8$ as well, the errors cannot be controlled then because of  the scarcity of NR simulations. But this issue doesn't prevent us from \emph{comparing} different models at higher spins. A different application of the balance-law could be to discriminate between choices made during modeling. For example, one could compare the consistency of different choices of extrapolation made by models outside the parameter range of NR. From the perspective of future detectors with significant enhancement in sensitivity, it is also important to reduce the main sources of error we encountered by including the spin-spin interaction terms in the calculation of $J^k(t_i)$ and using a more accurate method to evolve the  spins of the two black holes.

To summarize, we have shown that the angular momentum balance-law can be a valuable tool. It allows one compare models across all points in parameter space; enables one to identify---without the need of NR simulations---parameter ranges in which  errors are higher in specific models; provides guidance to waveform models for further improvements; informs us on the accuracy of the match between NR and PN parameters that are used to label the waveforms; and, even offers checks on the numerical simulations themselves.

\section*{Acknowledgments}
This work was supported by the NSF Grant No.~PHY-1806356, the Eberly Chair funds of Penn State, and the Mebus Fellowship to N.K. Computations for this research were performed on the Roar supercomputer of the Pennsylvania State University's   Institute for Computational and Data Sciences. We thank Frank Ohme and Angela Borchers Pascual for discussions and  comments; Eric Thrane for the suggestion of adding memory to the NR waveforms; and the referee for suggesting that we add the plots that now appear in Appendix B. We also acknowledge the use of the LAL Simulation \cite{lalsuite}, PyCBC \cite{pycbc}, scri \cite{scri} and spinsfast \cite{Huffenberger:2010hh} software packages in this paper.

\appendix
\begin{figure*}[htb]
        \centering
        \includegraphics{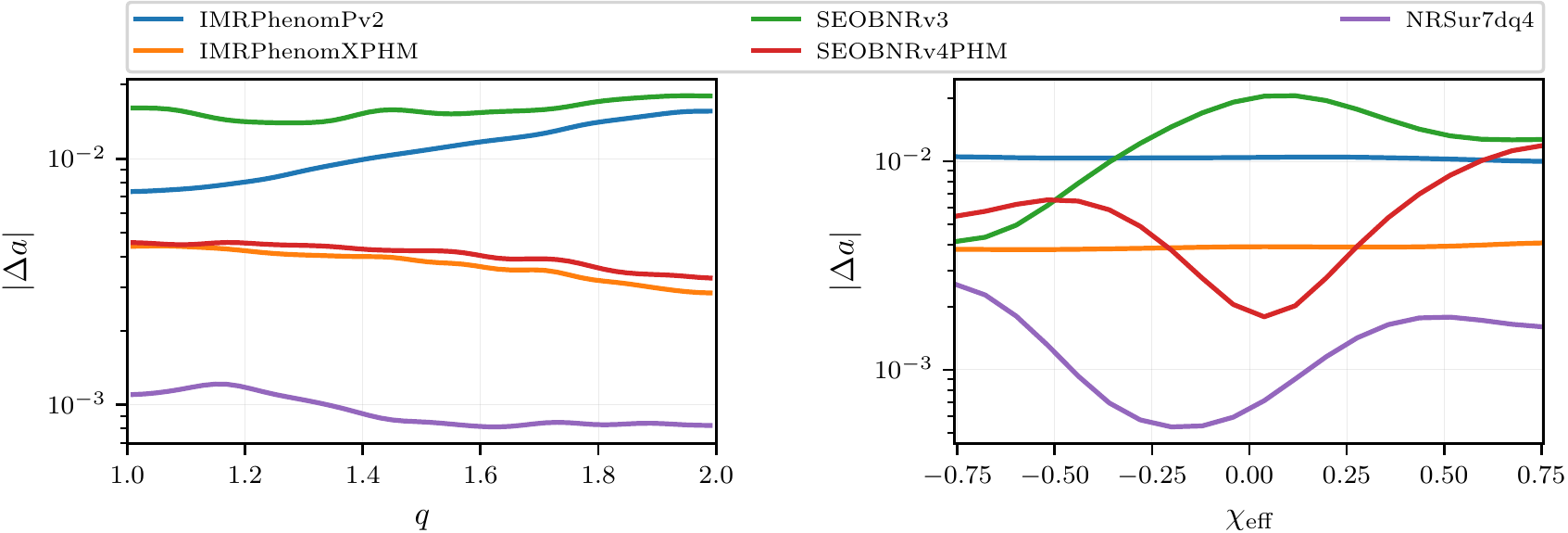}
        \caption{Nonprecessing systems: Distribution of balance-law violation for various models,  measured by $|\Delta a|$ as a function of mass ratio $q$ for the plot on the left, and $\chi_\mathrm{eff}$ for the plot on the right. The dependence of the error on other parameters is marginalized by averaging over bins. } 
        \label{fig:NP_qchi}
\end{figure*}

\begin{figure*}[htb]
 \centering
 {\Large $|\Delta a|$}
        \includegraphics{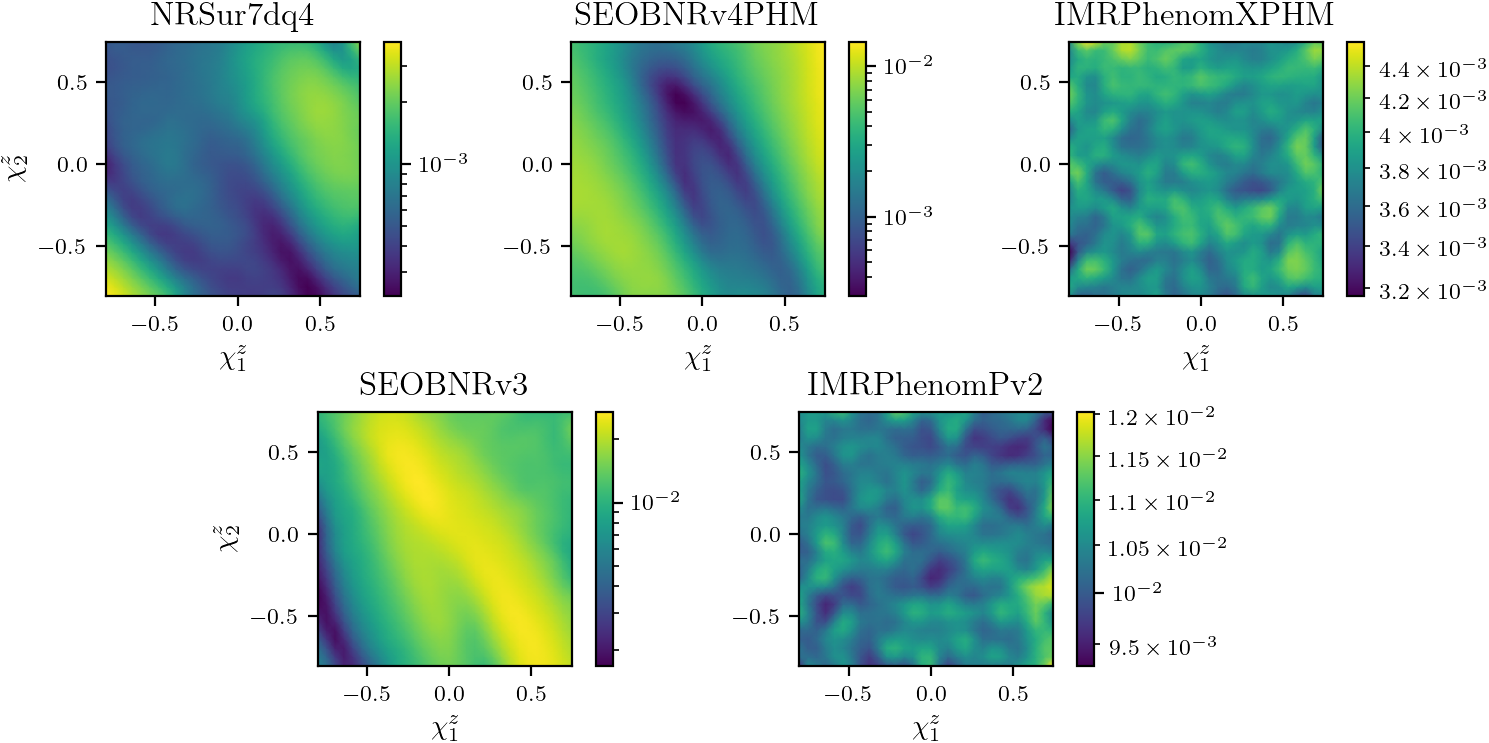}
        \caption{Nonprecessing systems: Distribution of balance-law violation measured by $|\Delta a|$ as a function of the spins of the black hole in the $z$-direction, $\chi_1^z$ and $\chi_2^z$. Note that by convention we have that $\chi_1^z$ is the spin of the heavier black hole.} 
        \label{fig:NP_chi_z}
\end{figure*}

\section{Post-Newtonian angular momentum}

\label{app:PN}
Here we describe in brief how to use the angular momentum formula from Eq.~(4.7) in \cite{Bohe:2012mr}, also reproduced below in Eq.~\eqref{eq:PN_ang_mom}, is used to get the initial angular momentum. To be consistent with \cite{Bohe:2012mr} we use boldface to denote vectors, $m=m_1+m_2$  to denote the total mass and restore factors of $G$ and $c$ in this section. The post-Newtonian formula is an expansion in the gauge invariant dimensionless PN parameter $x = (Gm\Omega_{\mathrm{orb}}/c^3)^{2/3}$, where $\Omega_\mathrm{orb}$ is the orbital angular frequency of the binary. We relate $\Omega_{\mathrm{orb}}$ at the beginning of the waveform to the starting frequency $f_{\mathrm{start}}$ of the $(2,2)$ mode of the  waveform model by $\Omega_{orb} = \pi f_\mathrm{start}$. Thus $x$ can be calculated from the starting frequency, and is $\approx 0.012$ for the dimensionless starting frequency of $5.8\times 10^{-3}$ used in this paper.

The conventions for the axes followed by the waveform models used in this paper are that at the reference time (taken to be the starting time of the waveform), the binary is separated along the $x$-axis, and instantaneously orbits counterclockwise in the $x,y$ plane. However for NR simulations  at the reference time---i.e. at a time when most of the junk radiation has passed through the outer boundary \cite{Boyle:2019kee}---the frame is arbitrary in general. Thus we perform a rotation to bring NR into the same frame conventions as the waveform models at its reference time. We solve for the rotation that brings the coordinate separation between the black holes be along the $x$-direction, and the angular velocity as defined in \cite{Boyle:2013nka} along the $z$-direction. Note that the separation of the black holes is in general a gauge dependent quantity. However it is still an essential ingredient of specifying the system that is integral to any comparison of the waveform against PN or a waveform model. Once we have fixed the frame, unit vectors along the $x,y,z$ directions at the start time are denoted $\mathbf{n},\boldsymbol\lambda, \boldsymbol\ell$ respectively. Note that this is only at the reference time and in general $\mathbf{n},\boldsymbol\lambda, \boldsymbol\ell$ evolve with time.

The spin variables convenient to use for the PN expressions are
\begin{align}
    \mathbf{S} = Gm_1^2\boldsymbol\chi_1 + Gm_2^2\boldsymbol\chi_2\,, \\
    \boldsymbol\Sigma = Gm m_2\boldsymbol\chi_2 - Gm m_1 \boldsymbol\chi_1\,.
\end{align}
The $x,y,z$ components of these vectors are $S_n, S_\lambda, S_\ell$ and $\Sigma_n, \Sigma_\lambda, \Sigma_\ell$.  It is also convenient to use the parameters total mass $m$, symmetric mass ratio $\nu= m_1m_2/(m_1+m_2)^2$ and $\delta m=m_1-m_2$. Finally, the total angular momentum $\mathbf{J}$ is given by
\begin{equation}
    \mathbf{J} = \mathbf{L} + \mathbf{S}/c\,, 
\end{equation}
where $\mathbf{L}$ is the orbital angular momentum and for quasi circular binaries. Keeping only terms in the 3.5 PN expansion that are linear in spin, $\mathbf{L}$ is given by \cite{Bohe:2012mr},
\begin{widetext}
\begin{align}
\label{eq:PN_ang_mom}
\mathbf{L}=\frac{G m^2 }{c\, x^{1/2}}\nu \Bigg\{ &
\boldsymbol\ell\left[
1
+x \left(\frac{3}{2} + \frac{1}{6} \nu\right)
+x^2 \left(\frac{27}{8} -\frac{19}{8} \nu + \frac{1}{24} \nu^2\right)\right.\nonumber\\
&\qquad\left.+x^3 \left(\frac{135}{16} +\left[- \frac{6889}{144} + \frac{41}{24} \pi^2\right] \nu +\frac{31}{24} \nu^2 + \frac{7}{1296} \nu^3 \right)\right] \nonumber\\
&+\frac{x^{3/2}}{G m^2}\Bigg(
\boldsymbol\ell\left[-\frac{35}{6}S_\ell-\frac{5}{2}\frac{\delta m}{m}\Sigma _\ell\right]
+\boldsymbol\lambda\left[-3S_{\lambda }-\frac{\delta m}{m}\Sigma _{\lambda }\right]
+\mathbf{n}\left[\frac{1}{2}S_n+\frac{1}{2}\frac{\delta m}{m}\Sigma _n\right]
\Bigg)\nonumber\\
&+\frac{x^{5/2}}{G m^2}\Bigg(
\boldsymbol\ell\left[\left(-\frac{77}{8} + \frac{427}{72} \nu\right)S_\ell+\frac{\delta m}{m}\left(-\frac{21}{8} + \frac{35}{12} \nu\right)\Sigma _\ell\right]\nonumber\\
&\qquad\qquad+\boldsymbol\lambda\left[\left(-\frac{7}{2} + 3 \nu\right)S_{\lambda }+\frac{\delta m}{m}\left(-\frac{1}{2} + \frac{4}{3} \nu\right)\Sigma _{\lambda }\right]\nonumber\\
&\qquad\qquad+\mathbf{n}\left[\left(\frac{11}{8} -\frac{19}{24} \nu\right)S_n+\frac{\delta m}{m}\left(\frac{11}{8} -\frac{5}{12} \nu\right)\Sigma _n\right]
\Bigg)\nonumber\\
&+\frac{x^{7/2}}{G m^2}\Bigg(
\boldsymbol\ell\left[\left(-\frac{405}{16} + \frac{1101}{16} \nu -\frac{29}{16} \nu^2\right)S_\ell+\frac{\delta m}{m}\left(-\frac{81}{16} + \frac{117}{4} \nu -\frac{15}{16} \nu^2\right)\Sigma _\ell\right]\nonumber\\
&\qquad\qquad
+\boldsymbol\lambda\left[\left(-\frac{29}{4} + \frac{1}{12} \nu -\frac{4}{3} \nu^2\right)S_{\lambda }+\frac{\delta m}{m}\left(-\frac{1}{2} -\frac{79}{24} \nu -\frac{2}{3} \nu^2\right)\Sigma _{\lambda }\right]\nonumber\\
&\qquad\qquad
+\mathbf{n}\left[\left(\frac{61}{16} -\frac{1331}{48} \nu + \frac{11}{48} \nu^2\right)S_n+\frac{\delta m}{m}\left(\frac{61}{16} -\frac{367}{24} \nu + \frac{5}{48} \nu^2\right)\Sigma _n\right]
\Bigg)\nonumber\\
&+\mathcal{O}\left(\frac{1}{c^8}\right)
\Bigg\} \;.
\end{align}
\end{widetext}
 Thus, given the masses and spins of the two black holes in the binary and the initial orbital frequency, one can use Eq.~\eqref{eq:PN_ang_mom} to obtain the initial angular momentum.

\begin{figure*}[htb]
\centering
{\Large $|\Delta a|$}
\includegraphics{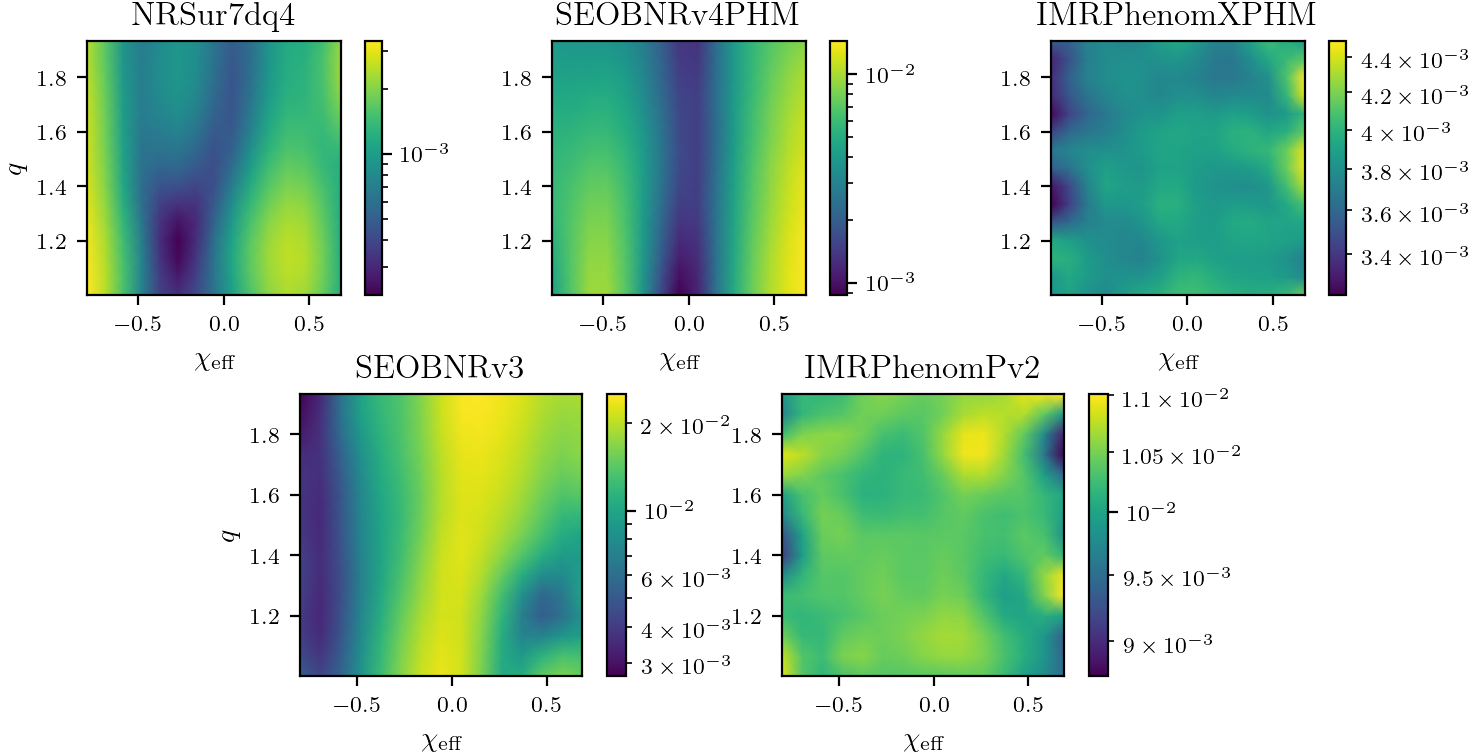}
\caption{Nonprecessing systems: Distribution of of balance-law violation measured by $|\Delta a|$ as a function of mass ratio $q$ and $\chi_\mathrm{eff}$.}
\label{fig:NP_q_chieff}
\end{figure*}

\begin{figure*}[htb]
\centering
        \includegraphics{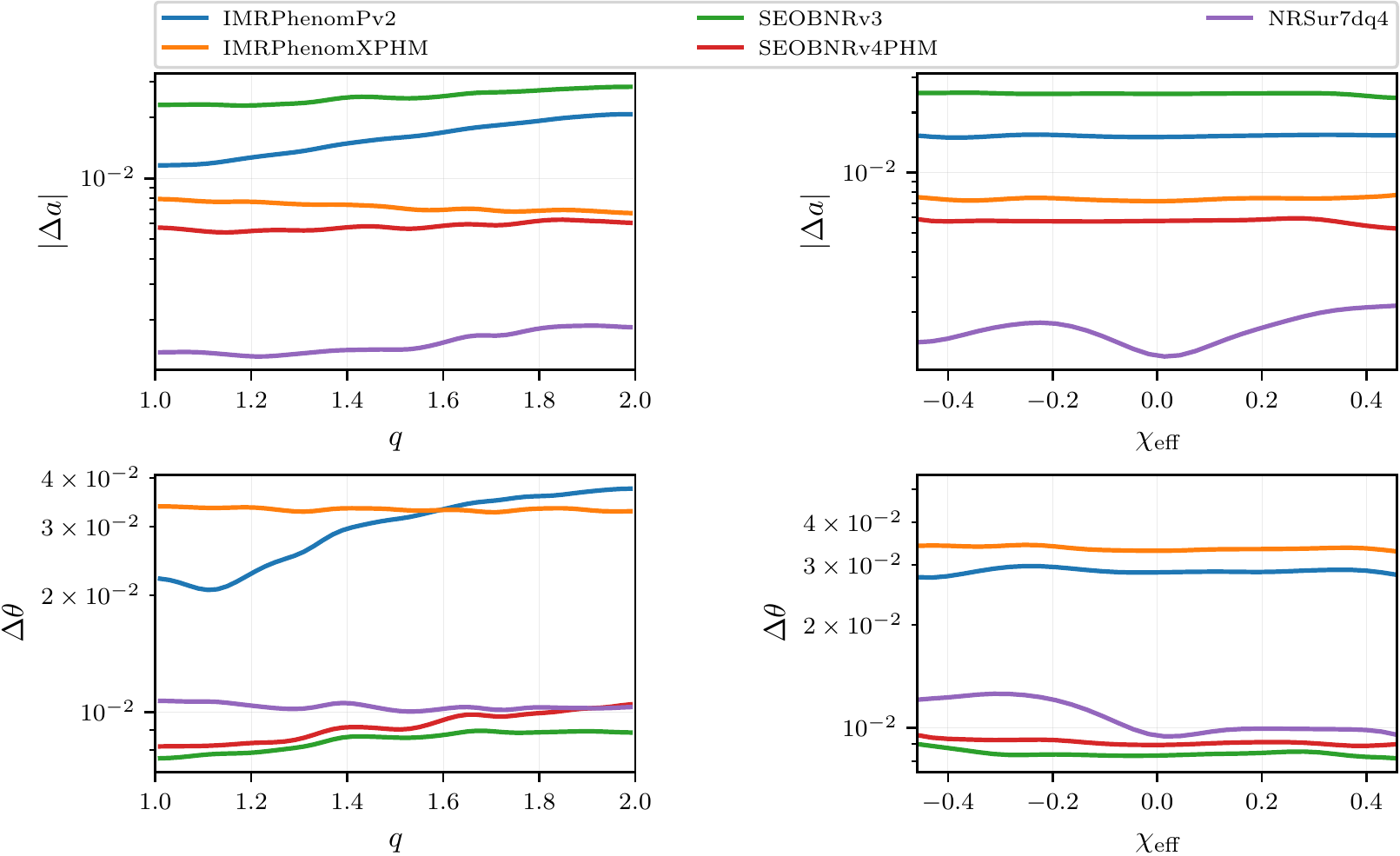}
        \caption{Precessing systems: The distribution of balance-law violation as a function of mass ratio $q$ and $\chi_\mathrm{eff}$. The upper plots show the violation measured by the magnitude $|\Delta a|$ and the lower plots show the violation measured by the angle $\Delta\theta$.}
        \label{fig:P_qchi}
\end{figure*}

\begin{figure*}[htb]
\centering
 {\Large $|\Delta a|$}
\includegraphics{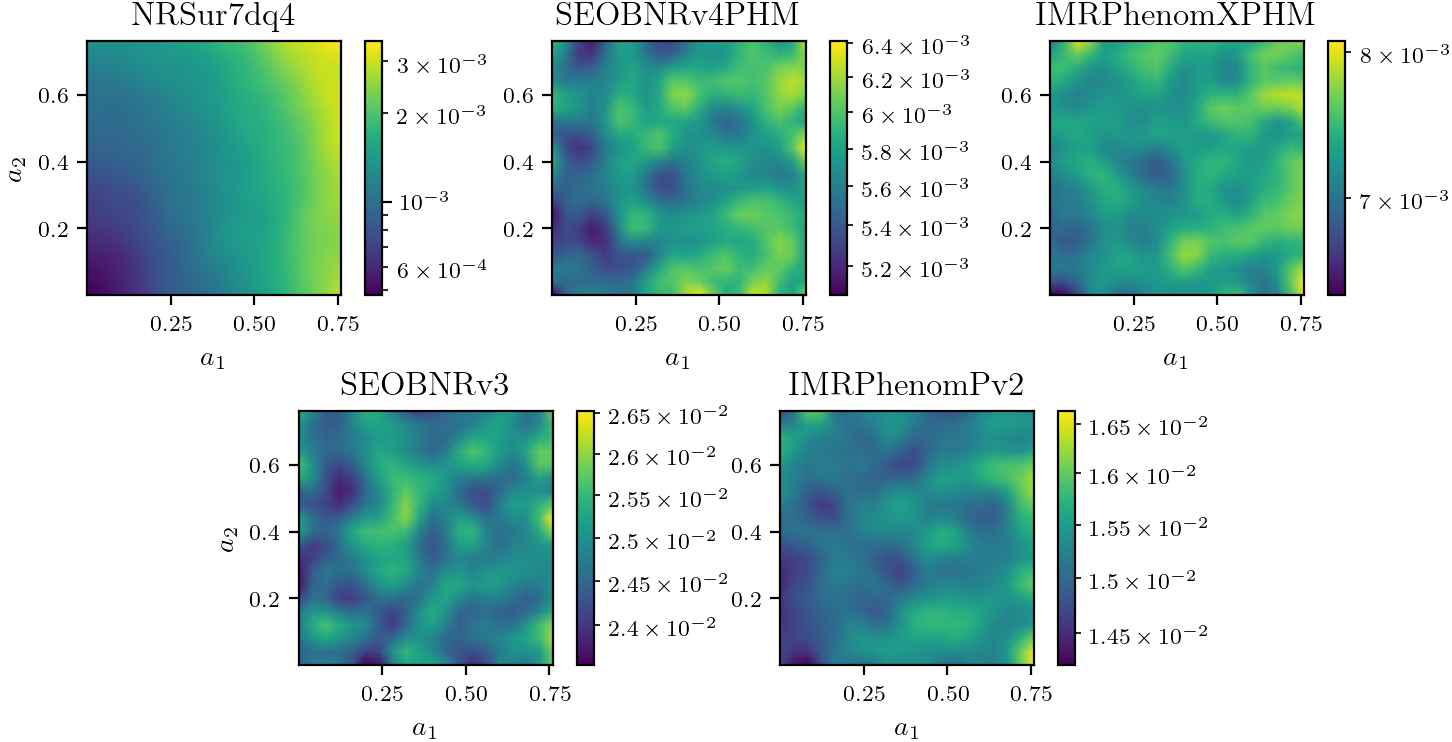}
\caption{Precessing systems: The distribution of the violation in magnitude $|\Delta a|$ of the balance-law as a function of the black hole spin magnitudes $a_1$ and $a_2$.}
\label{fig:P_chinorm_mag}
\end{figure*}

\begin{figure*}[htb]
\centering
 {\Large $|\Delta \theta|$}
\includegraphics[width=\textwidth]{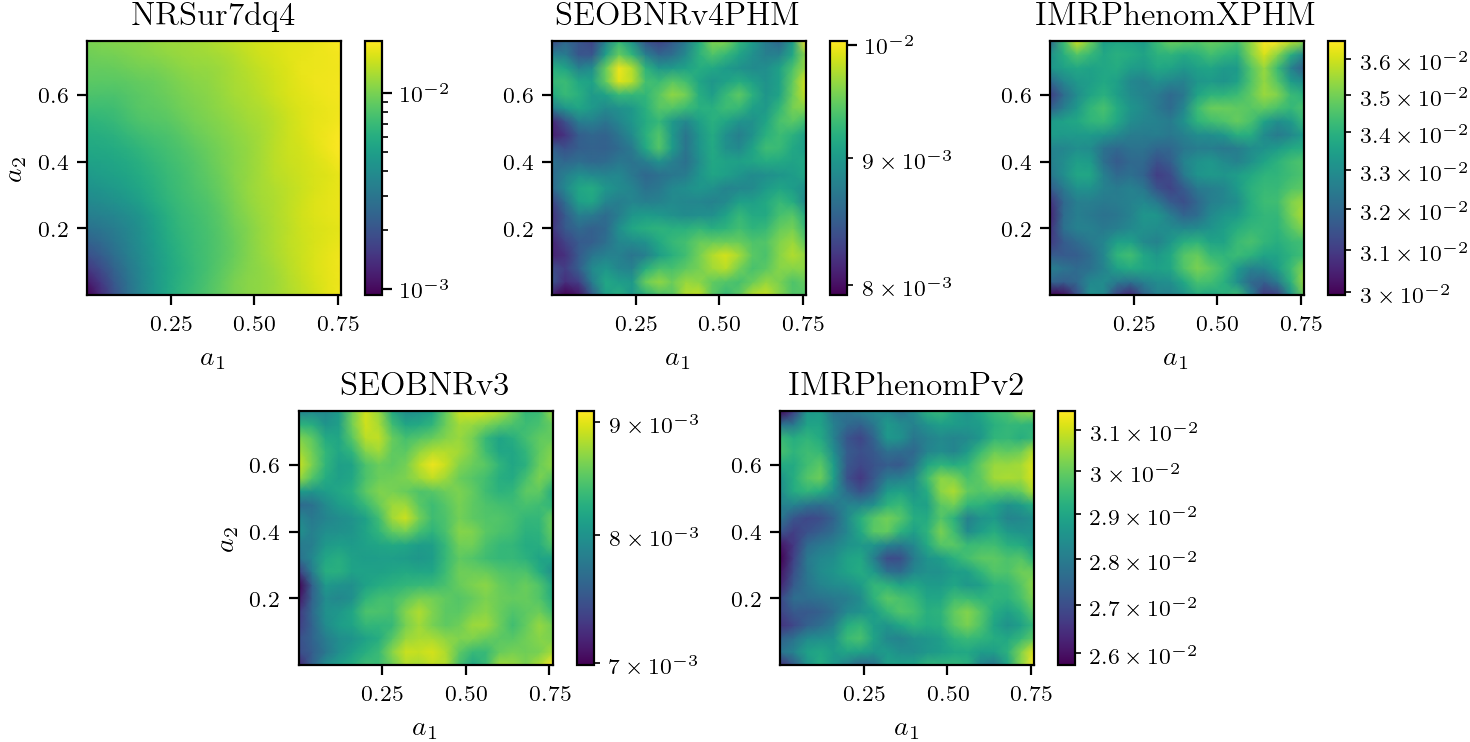}
\caption{Precessing systems: The distribution of the angle $|\Delta\theta|$ of the balance-law violation as a function of the black hole spin magnitudes $a_1$ and $a_2$.}
\label{fig:P_chinorm_angle}
\end{figure*}

\begin{figure*}[htb]
 {\Large $|\Delta a|$}
\includegraphics{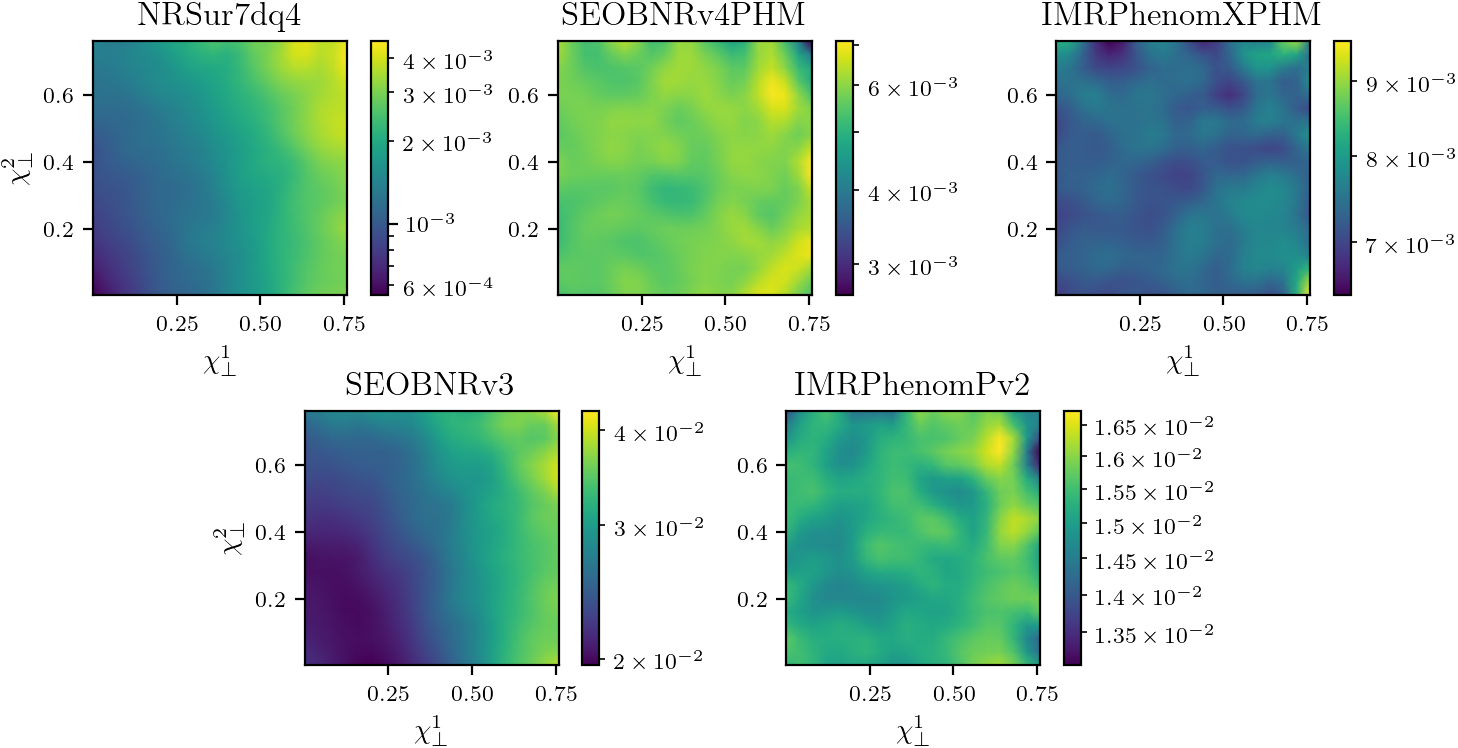}
\caption{Precessing systems: The distribution of the violation in magnitude $|\Delta a|$ of the balance-law violation as a function of the black hole spin magnitudes in the perpendicular direction $\chi_\perp^1$ and $\chi_\perp^2$.}
\label{fig:P_chip_mag}
\end{figure*}

\begin{figure*}[htb]
 {\Large $|\Delta \theta|$}
\includegraphics{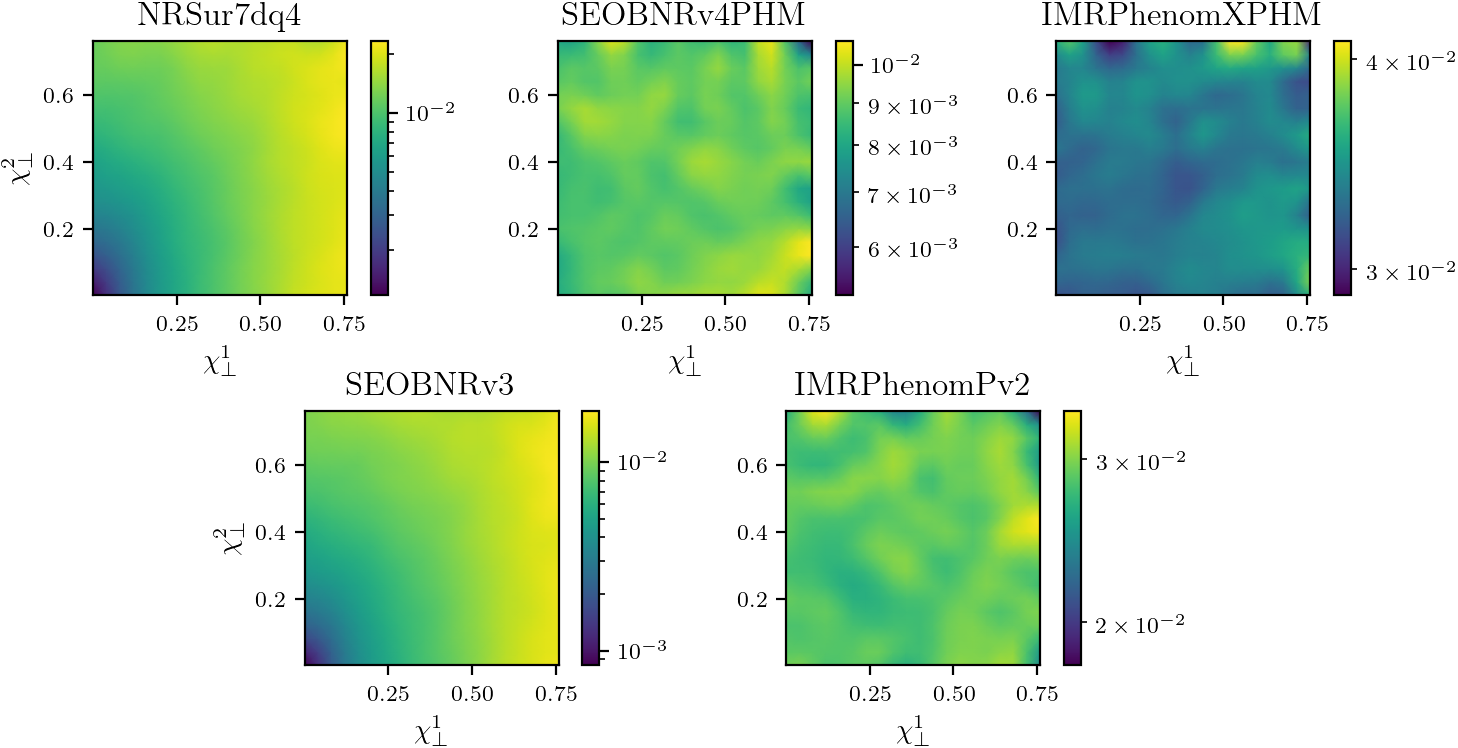}
\caption{Precessing systems: The distribution of the angle $|\Delta\theta|$ of the balance-law violation as a function of the black hole spin magnitudes in the perpendicular direction $\chi_{\perp}^1$ and $\chi_{\perp}^2$.}
\label{fig:P_chip_angle}
\end{figure*}

\begin{figure*}[htb]
{\large $|\Delta a|$}
\includegraphics{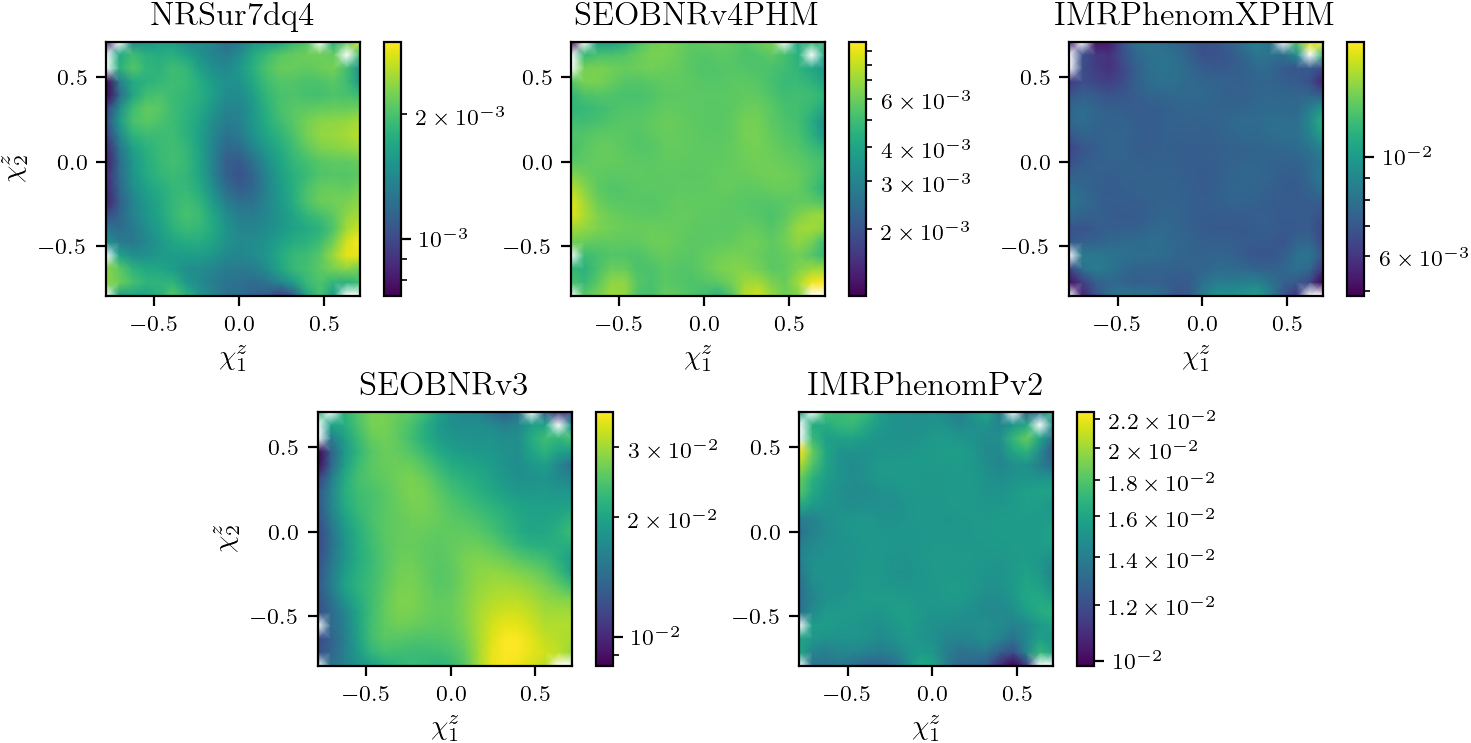}
\caption{Precessing systems: The distribution of the violation in magnitude $|\Delta a|$ of the balance-law as a function of the black hole spins in the $z$ direction $\chi_{1}^z$ and $\chi_{2}^z$. Because for an isotropic distribution of spins, $\chi_z$ being close to the maximum value is suppressed, the corners do not have sufficient points; bins without any points are in white.}
\label{fig:P_chiz_mag}
\end{figure*}

\begin{figure*}[htb]
{\large $|\Delta \theta|$}
\includegraphics{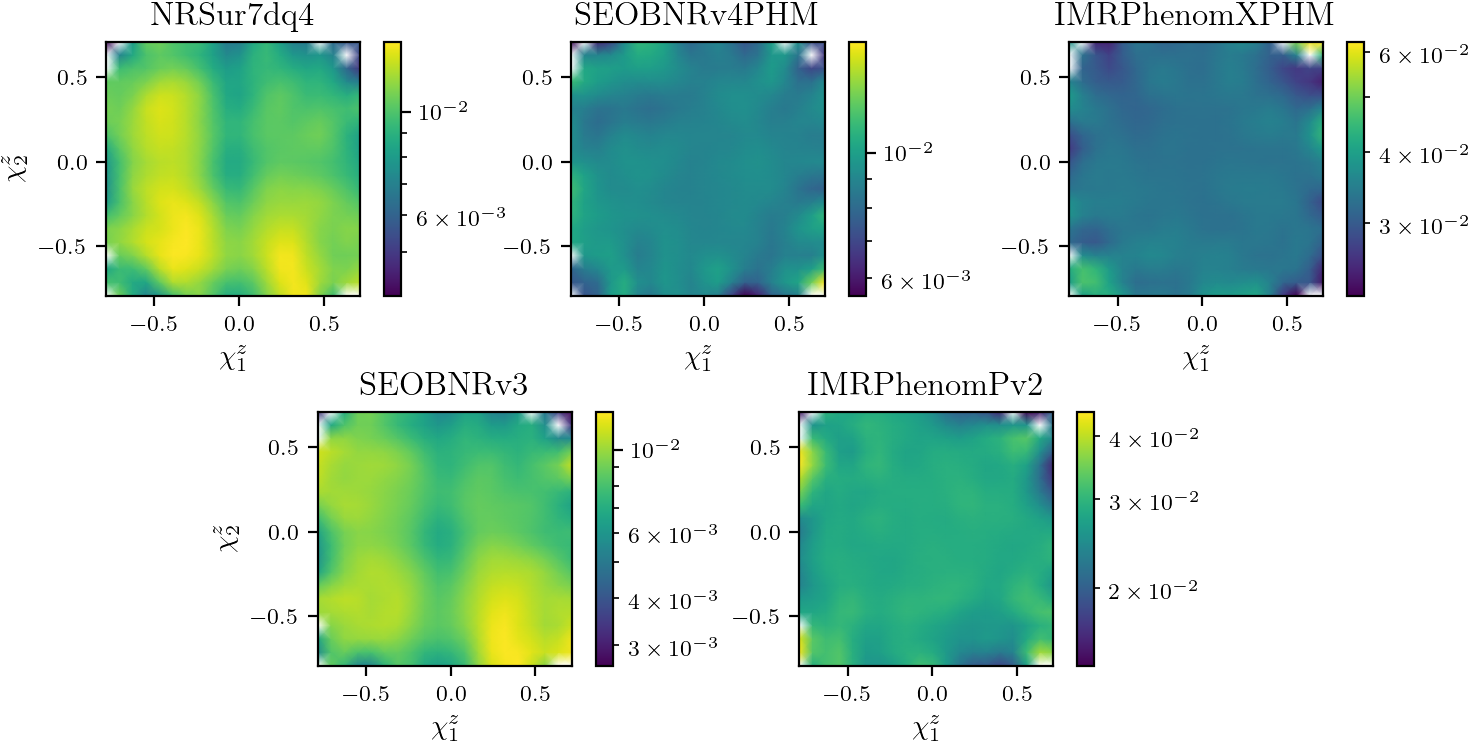}
\caption{Precessing systems:The distribution of the angle $|\Delta \theta|$ of the balance-law violation as a function of the black hole spins in the $z$ direction $\chi_{1}^z$ and $\chi_{2}^z$. Because for an isotropic distribution of spins, $\chi_z$ being close to the maximum value is supressed, the corners do not have sufficient points; bins without any points are in white.}
\label{fig:P_chiz_angle}
\end{figure*}

\section{Balance-law violation as function of Parameter space}
\label{app:Heat_map}

As discussed, the balance-law can be applied to any point in parameter space and this can help identify regions with higher errors. These results can help provide pointers to where newer models can be improved, and where more numerical simulations are needed. In this appendix we present various results exploring the parameter space for the different models considered in this paper. and identify such regions of parameter space for the different models.  As before we divide the study into the family of precessing systems and nonprecessing systems, and use the same distribution of points as in Sec.~\ref{sec:Results}. Furthermore as before, for precessing systems we quantify the error in the magnitude $|\Delta a| = |a_\mathrm{bal}-a_{fit}|$ and in angle $\Delta\theta$ between $\vec{\chi}_\mathrm{bal}$ and $\vec{\chi}_\mathrm{fit}$ separately. In Fig.~\ref{fig:NP_qchi} and Fig.~\ref{fig:P_qchi} we show the error dependence on mass ratio $q$ and $\chi_\mathrm{eff}$. Since the error depends on many more parameters, we marginalize over them by binning, followed by smoothing by a Gaussian convolution. The mass ratio $q$ does not change strongly with parameter space, however we see several features of the violation in terms of $\chi_\mathrm{eff}$. For the nonprecessing systems we study the spin dependence of the violation in  Fig.~\ref{fig:NP_chi_z}. We again use binning to marginalize over other parameters, followed by a smoothing. We find that the Phenom models' balance-law violation varies weakly with the spins. For EOB and surrogate models the figure identifies regions of spin-space where errors are larger and smaller. We also see in Fig.~\ref{fig:NP_q_chieff} that, for nonprecessing systems, while \texttt{NRSur7dq4} has a slight correlation of $|\Delta a|$ between $q$ and $\chi_\mathrm{eff}$, for the other models these parameters are very uncorrelated. 

For precessing systems the spin-space has more dimensions, and thus more choices of spin variables are possible. In Fig.~\ref{fig:P_chinorm_mag} and Fig.~\ref{fig:P_chinorm_angle} we show error dependence on the magnitudes of the spins, in Fig.~\ref{fig:P_chip_mag} and Fig.~\ref{fig:P_chip_angle} in terms of the spin projected on the orbital plane, $\chi_\perp$, and finally in Fig.~\ref{fig:P_chiz_mag} and Fig.~\ref{fig:P_chiz_angle} in terms of the $z$ components of the spins. These figures highlight several differences between the models. Overall for the Phenom models we do not see large variations of the violation across parameter space. For the surrogate models we see a trend of increasing violations with higher spins. Finally while \texttt{SEOBNRv3} has several features in these figures we do not see any distinctive features for \texttt{SEOBNRv4PHM} for the precessing systems.  

\clearpage

    \bibliographystyle{apsrev4-1}
    \bibliography{references}

\end{document}